\documentstyle[10pt,aaspp4]{article}
\def\ngc#1{\hbox{NGC$\,$#1}}

\def\redn1365{0.16 $\pm$ 0.08 mag\,}
\def\distmod{31.31 $\pm$ 0.08 mag\,}

\def\distn1365{18.3 $\pm$ 0.7 Mpc\,}

\def\av{0.40\,}
\def\muv{31.70 $\pm$ 0.05 mag\,}
\def\mui{31.54 $\pm$ 0.06 mag\,}
\def\domuv{31.64 $\pm$ 0.07 mag\,}
\def\domui{31.49 $\pm$ 0.07 mag\,}
\def\domod{31.26 $\pm$ 0.10 mag\,}
\begin{document}
\title{The HST Key Project on the Extragalactic Distance Scale XIV. \\
The Cepheids in \ngc{1365}\footnote{Based on observations with the NASA/ESA
{\it Hubble Space Telescope} obtained at the Space Telescope Science
Institute, which is operated by AURA, Inc. under NASA Contract No.
NAS5-26555.}}

\author{N. A. Silbermann,\footnote{Infrared Processing and Analysis Center,
Jet Propulsion Laboratory, California Institute of Technology,
MS 100-22, Pasadena, CA 91125} 
Paul Harding,\footnote{Steward Observatory,
University of Arizona, Tucson, AZ 85721}
Laura Ferrarese,\footnote{Hubble Fellow, California Institute of 
Technology, Pasadena, CA 91125}  
Peter B. Stetson,\footnote{Dominion Astrophysical Observatory,
Victoria, British Columbia V8X 4M6 Canada}
Barry F. Madore,\footnote{NASA Extragalactic Database,
Infrared Processing and Analysis
Center, California Institute of Technology, MS 100-22, Pasadena, CA 91125}
\addtocounter{footnote}{-4}
Robert C. Kennicutt, Jr.,\footnotemark
\addtocounter{footnote}{3}
Wendy L. Freedman,\footnote{Observatories of the Carnegie
Institution of Washington, Pasadena CA 91101}
Jeremy R. Mould,\footnote{Mount Stromlo and Siding Spring
Observatories, Institute of Advanced Studies, ANU, ACT 2611 Australia}
\addtocounter{footnote}{-6}
Fabio Bresolin,\footnotemark
\addtocounter{footnote}{5}
Holland Ford,\footnote{Johns Hopkins University and Space Telescope Science
Institute, Baltimore, MD 21218}
\addtocounter{footnote}{-2}
Brad K. Gibson,\footnotemark
\addtocounter{footnote}{1}
John A. Graham,\footnote{Dept. of Terrestrial Magnetism,
Carnegie Institution of Washington, Washington D.C. 20015}
Mingsheng Han,\footnote{University of Wisconsin,
Madison, Wisconsin 53706}
\addtocounter{footnote}{-1}
John G. Hoessel,\footnotemark
Robert J. Hill,\footnote{Laboratory for Astronomy
\& Solar Physics, NASA Goddard Space
Flight Center, Greenbelt MD 20771}
John Huchra,\footnote{Harvard Smithsonian
Center for Astrophysics, Cambridge, MA 02138}
Shaun M.G. Hughes,\footnote{Royal Greenwich
Observatory, Cambridge CB3 0EZ England}
Garth D. Illingworth,\footnote{Lick Observatory,
University of California, Santa Cruz CA 95064}
\addtocounter{footnote}{-1}
Dan Kelson,\footnotemark
\addtocounter{footnote}{-3}
Lucas Macri,\footnotemark
\addtocounter{footnote}{-7}
Randy Phelps,\footnotemark
Daya Rawson,\footnotemark
\addtocounter{footnote}{-7}
Shoko Sakai,\footnotemark
and Anne Turner\footnotemark
}

\begin{abstract}

We report the detection of Cepheid variable stars in the
barred spiral galaxy \ngc{1365}, located in the Fornax cluster,
using the Hubble Space Telescope Wide
Field and Planetary Camera 2.  Twelve $V$ (F555W) and four $I$
(F814W) epochs of observation were obtained. 
The two photometry packages, ALLFRAME and DoPHOT, were
separately used to obtain profile-fitting photometry of all the stars
in the HST field. The search for Cepheid variable stars resulted in
a sample of 52 variables, with periods between 14 and 60 days,
in common with both datasets.
ALLFRAME photometry and light curves of the Cepheids are presented.
A subset of 34 Cepheids were selected on the basis of period, 
light curve shape, similar
ALLFRAME and DoPHOT periods, color, and relative crowding,
to fit the Cepheid period-luminosity relations in $V$ and $I$ for
both ALLFRAME and DoPHOT.
The measured distance modulus to \ngc{1365} from the ALLFRAME photometry
is 31.31 $\pm$ 0.20 (random) $\pm$ 0.18 (systematic) mag,
corresponding to a distance of 
18.3 $\pm$ 1.7 (random) $\pm$ 1.6 (systematic) Mpc. 
The reddening is measured to be $E(\hbox{\it V--I)}$ = \redn1365. 
These values are in excellent agreement
with those obtained using the DoPHOT photometry, namely a distance
modulus of 31.26 $\pm$ 0.10 mag, and a reddening of 0.15 $\pm$ 0.10 mag
(internal errors only). 
\end{abstract}


\section{Introduction}

The observations presented in this paper are part of the 
{\sl Hubble Space Telescope} Key Project on the Extragalactic Distance Scale,
a detailed description of which can be found in
\markcite{kfm1995}Kennicutt, Freedman \& Mould (1995). 
The main goal of the Key Project
is to measure the Hubble Constant, H$_{0}$, to an accuracy of 10\%, by
using Cepheids to calibrate several secondary distance indicators such as
the planetary nebula luminosity
function, the Tully--Fisher relation, surface brightness fluctuations,
and methods using supernovae.
These methods will then be used to
determine the distances to more distant galaxies whereby the global
Hubble Constant can be measured.
Published results from other Key Project galaxies
include M81 (\markcite{free1994}Freedman {\it et al.} 1994), 
M100 (\markcite{lff1996}Ferrarese {\it et al.} 1996),
M101 (\markcite{kelson1996}Kelson {\it et al.} 1996), 
\ngc{925} (\markcite{silb1996}Silbermann {\it et al.} 1996),
\ngc{3351} (\markcite{grah1997}Graham {\it et al.} 1997), 
\ngc{3621} (\markcite{raw1997}Rawson {\it et al.} 1997),
\ngc{2090} (\markcite{phel1998}Phelps {\it et al.} 1998),
\ngc{4414} (\markcite{turn1998}Turner {\it et al.} 1998),
\ngc{7331} (\markcite{hughes1998}Hughes {\it et al.} 1998), and
\ngc{2541} (\markcite{lff1998a}Ferrarese {\it et al.} 1998).
 
NGC~1365 ($\alpha_{1950} = 3^{\rm h} 31^{\rm m}$, $\delta_{1950} = -36\arcdeg
18\arcmin$) is a large, symmetric, barred spiral galaxy with a measured
heliocentric velocity of 1652 km~s$^{-1}$ \markcite{st1981} 
(Sandage \& Tammann 1981) located in the Fornax cluster of galaxies.
It is classified as an SBb(s)I galaxy by \markcite{st1981}
Sandage \& Tammann and as an SBs(b) galaxy by
\markcite{dev1991}de Vaucouleurs {\it et al.} (1991). 
Work by Veron {\it et al.} (1980) found that \ngc{1365} contains a
hidden Seyfert 1 nucleus. \ngc{1365} has been extensively mapped in HI
by Ondrechen \& van der Hulst (1989) and more recently by
\markcite{jvm1995}J\"{o}rs\"{a}ter \& van Moorsel (1995). In particular,
\markcite{jvm1995}J\"{o}rs\"{a}ter \& van Moorsel found that the 
inner disk of \ngc{1365} has a significantly different inclination
angle (40\arcdeg) compared to previous work using optical isophotes (55\arcdeg)
by Linblad (1978). Other inclination angles found in the literature are
56\arcdeg (Bartunov {\it et al.} 1994), 61\arcdeg (Schoniger \& Sofue 1994,
Aaronson {\it et al.} 1981),
44\arcdeg~$\pm$ 5\arcdeg~(Bureau, Mould, \& Staveley-Smith 1996),
46\arcdeg $\pm$ 8\arcdeg~(Ondrechen \& van der Hulst 1989)
and 63\arcdeg~(Tully 1988). This scatter is a result of 
the warped nature of the \ngc{1365} disk
(\markcite{jvm1995}J\"{o}rs\"{a}ter \& van Moorsel 1995) combined with the 
various observational methods the authors used to determine the
the major and minor axis diameters.
The true inclination angle of \ngc{1365} is not vital for our Cepheid
work but is critial for photometric and HI line-width corrections
for the Tully-Fisher (TF) relation. If one uses infrared ($H$ band) absolute 
magnitudes the effect of using the various inclination angles between
40\arcdeg~to 63\arcdeg~is minimal as the extinction correction in the infrared
is small. 
However, the correction to the line-width, (sin $i$)$^{-1}$, is not small.
The corrected HI line width will decrease by
40\% if one uses 63\arcdeg~instead of 40\arcdeg. Detailed discussion of
the TF method is beyond the scope of this paper but we caution readers
to accurately determine the inclination angle of \ngc{1365} before using
it as a TF calibrator.

Overall, Fornax is a relatively compact cluster of about 350 galaxies
with a central, dense concentration of elliptical galaxies.
The center of the cluster is dominated by the large E0 galaxy
\ngc{1399}, with \ngc{1365} lying $\sim$0.5 Mpc from \ngc{1399} as projected
on the sky. The heliocentric radial velocity of the Fornax cluster is 
1450 km~s$^{-1}$ with a dispersion
of 330 km~s$^{-1}$ (\markcite{hm1994}Held \& Mould 1994). 
The Fornax
cluster has been the target of many studies involving secondary distance 
indicators, as reviewed by Bureau, Mould, \& Staveley-Smith (1996). 
A Cepheid distance to the cluster will be an important step
forward in calibrating the extragalactic distance scale.

This is the first of two papers on the Cepheid distance to \ngc{1365}
and the Fornax cluster. This paper describes the HST observations of
Cepheid variable stars in \ngc{1365} and the distance derived to the
galaxy. A companion paper by \markcite{mad1998a} Madore {\it et al.} (1998a) 
discusses the
implications for the distance to the Fornax cluster and the calibration
of the extragalactic distance scale.
The location of \ngc{1365} within the Fornax cluster and the geometry
of the local universe is discussed in \markcite{mad1998b}Madore {\it et al.}
(1998b).
We note that the Key Project has recently observed two other galaxies within 
the Fornax cluster, \ngc{1425} (Mould {\it et al.} 1998) and \ngc{1326A} 
(Prosser {\it et al.} 1998).

\section{Observations}

NGC~1365 was imaged using the Hubble Space Telescope's Wide Field and
Planetary Camera 2 (WFPC2). A description of
WFPC2 instrument is given in the HST WFPC2 Instrument Handbook
(\markcite{burr1994}Burrows {\it et al.} 1994).  The camera
consists of four 800$\times$800 pixel CCDs.
Chip 1 is the Planetary Camera with 0.046 arcsec pixels and an illuminated
$\sim 33\times31$ arcsec field of view. The three other CCDs make up
the Wide Field Camera (Chips 2-4), each with 0.10 arcsec pixels and a
$\sim 1.25\times1.25$ arcmin illuminated field of view. Each CCD has a readout
noise of about 7 $e^{-}$. A gain setting of 7 $e^{-}$/ADU was used for all
of the \ngc{1365} observations.

WFPC2 imaged the eastern part of \ngc{1365},
as seen in Figure 1, from 1995 August through September. As with all the 
galaxies chosen for the Key Project,
the dates of observation were selected using a power-law time series
to minimize period aliasing and maximize uniformity of phase coverage
for the expected range of Cepheid periods from 10 to 60 days
(\markcite{free1994}Freedman {\it et al.} 1994).  Twelve epochs in
$V$ (F555W) and four in $I$ (F814W) were obtained. Each 
epoch consisted of two exposures, taken on successive orbits,
with typical integration times of 2700 seconds. All observations
were made with the camera at an operating temperature of
$-$88$\arcdeg$ C.  Table 1 lists the image identification code,
Heliocentric Julian Date of observation (mid-exposure), exposure time, 
and filter, of each observation. On 1995 August 28 the focus of HST
was changed, occurring between the 5th and 6th epoch of \ngc{1365}
observations. The effect of this refocus is discussed in Section 3.1.

\section{Photometric Reductions}

All observations were preprocessed through the standard Space
Telescope Science Institute (STScI) pipeline as described by
\markcite{holt1995b}Holtzman {\it et al.}  (1995b).  The images were
calibrated with the most up-to-date version of the routine reference
files provided by the Institute at the time the images were taken.
Our post-pipeline processing included masking out the vignetted edges, 
bad columns, and bad pixels.
The images were then multiplied by a pixel area map to correct for the
WFPC2 geometric distortion (\markcite{hill1998}Hill {\it et al.} 1998), 
multiplied by 4, and converted to integer values. Then the ALLFRAME and
DoPHOT photometry packages were used to obtain profile-fitting photometry
of all the stars in the HST images.
Details of each method are described below.

\subsection{DAOPHOT II/ALLFRAME Photometry}

The extraction of stellar photometry from CCD images using the
ALLFRAME (\markcite{pbs1994}Stetson 1994) package first requires a 
robust list of stars in the
images. The long exposures of \ngc{1365} contain significant numbers of 
cosmic ray events which can be misidentified as stars by automated
star-finding programs. To solve this problem, images
for each chip were median averaged to produce a
clean cosmic ray free image. DAOPHOT II and ALLSTAR
(\markcite{pbs1987}Stetson 1987, \markcite{sdc1990}Stetson {\it et al.} 1990, 
Stetson 1991, 1992, 1994) were then used to identify the stars
in the deep, cosmic ray free images for each chip.
The star lists were subsequently used by ALLFRAME to
obtain profile-fitting photometry of the stars in the original
images. The point spread functions were derived from
public domain HST WFPC2 observations of the globular clusters
Pal 4, and \ngc{2419}.

Ten to twenty fairly isolated stars in each WFPC2 chip were chosen to
correct the profile-fitting ALLFRAME photometry out to the 0.5 arcsec
system of \markcite{holt1995a}Holtzman {\it et al.} (1995a). These
stars are listed in the appendix as secondary standards to our photometric
reductions.
To remove the effects of nearby neighbors on the aperture photometry
all the other stars were first subtracted from the images. Growth curves for
the isolated stars were constructed using DAOGROW 
(\markcite{pbs1990}Stetson 1990). 
Best estimates
of the aperture corrections were determined by running the program COLLECT
(\markcite{pbs1993}Stetson 1993), which uses the profile-fitting photometry
and the growth-curve photometry to determine the best photometric correction 
out to the largest aperture, 0.5 arcsec in our case. 

As with the previous Key Project galaxy \ngc{925} 
(\markcite{silb1996}Silbermann {\it et al.} 1996),
the aperture corrections (ACs) for all the frames in each filter were averaged 
to produce mean ACs that were then applied to each frame to shift the
Cepheid photometry onto the \markcite{holt1995a}Holtzman {\it et al.} 
0.5 arcsec system.
For nonvariable stars, mean instrumental $V$ and $I$ magnitudes averaged over
all epochs were calculated using DAOMASTER (\markcite{pbs1993}Stetson 1993) 
and then shifted to 
the 0.5 arcsec system using the mean ACs. However, for \ngc{1365} we
noticed there were relatively large variations in the individual frame
ACs, typically on the order of $\pm$ 0.07 mag, but for one case as large as
$-$0.2 mag, from the mean values. 
These variations are due to random photometric scatter in the small number of
isolated, bright stars available to us for AC determination, and the
realization that on any given image, a few of these stars may be corrupted by
comic ray events or chip defects further reducing the number of usable AC
stars for a particular image.
There was also a focus change between the
5th and 6th epoch, which resulted in an obvious shift of about +0.1 mag in
the individual ACs for epochs obtained after the focus change.
As a result, we also calculated mean magnitudes and fit period-luminosity 
relations for the Cepheids using photometry corrected by individual frame ACs.
The typical difference between
mean $V$ and $I$ magnitudes for the Cepheids (mean ACs $-$ individual ACs) 
was +0.01 mag, and for some
of the Cepheids was as large as +0.02 mag.
Due to the refocusing event and the relatively large variation in the 
individual ACs we feel that the individual photometric measurements of the
Cepheids at each epoch are best represented using the individual ACs. The
effect on the distance modulus to \ngc{1365} is minimal, +0.01 mag.
Throughout the rest of this paper we will
use the individual AC corrected Cepheid photometry.

The final form of the conversion equations for the ALLFRAME photometry is: 

\begin{equation}
M = m + 2.5\log t - C_{1}(V-I) + C_{2}(V-I)^{2} + AC + ZP
\end{equation}

\noindent 
where M the standard magnitude, m is the 
instrumental magnitude, $t$ is the exposure time, $C_{1}$ and $C_{2}$ are
the color coefficients, AC is the aperture 
correction (different for each frame for the Cepheids), and ZP is the zero 
point.
Since we have shifted our photometry to the 0.5 arcsec system of
\markcite{holt1995a}Holtzman {\it et al.} (1995a) we use their color
coefficients from their Table 7: $C_{1} = -0.052$ for $V$ and $-$0.063 for $I$,
and $C_{2} = 0.027$ for $V$ and 0.025 for $I$.
The ZP term includes the \markcite{hill1998}Hill {\it et al.} (1998) long 
exposure zero point, 
the ALLFRAME zero point of $-$25.0 mag,  and a correction for multiplying the 
images by four before converting them to integers (2.5log(4.0)).  For $V$ band 
the ZP terms are $-$0.984 $\pm$ 0.02, $-$0.973 $\pm$ 0.01, 
$-$0.965 $\pm$ 0.01, and $-$0.989 $\pm$ 0.02 for Chips 1$-$4 respectively,
and for
$I$ band are $-$1.879 $\pm$ 0.04, $-$1.838 $\pm$ 0.02, $-$1.857 $\pm$ 0.02, 
and $-$1.886 $\pm$ 0.01 for Chips 1$-$4 respectively.
For each epoch, an initial guess of each star's color, $V-I$ = 0.0, was used
and the $V$ and $I$ equations were then solved iteratively. 

For the nonvariable stars, an average magnitude over all epochs was calculated
using DAOMASTER, which corrects for frame-by-frame differences due to 
varying exposure times and focus, and the equations above were used to 
calculate the final standard $V$ and $I$ magnitudes using mean aperture 
corrections for each chip and filter combination. 

\subsection{DoPHOT Photometry}

As a double-blind check on our reduction procedures, we
separately reduced the \ngc{1365} data using a variation 
of DoPHOT (\markcite{schec1993}Schechter et al. 1993) described
by \markcite{saha1996}Saha {\it et al.} (1996).
The DoPHOT reductions followed the procedure described by
\markcite{saha1996}Saha {\it et al.},
\markcite{lff1996} Ferrarese {\it et al.} (1996) and 
\markcite{silb1996}Silbermann {\it et al.} (1996). Here we 
only mention aspects of the reductions that were unique to NGC 1365.
The DoPHOT reduction procedure identifies cosmic rays when
combining the single epoch exposures, prior to running DoPHOT
(\markcite{saha1994}Saha {\it et al.} 1994).  
Due to the unusually long exposure times in
the \ngc{1365} images, however, the number of pixels affected by cosmic
rays was so large that significant
numbers of residual cosmic ray artifacts remained in the combined
images.  To overcome this problem, the combined images for each epoch
were further combined in pairs, to create a single clean master image.
The master image was then compared with each original image to identify
and flag the cosmic rays.  The process was then repeated, but with
the pixels affected by cosmic ray events in both single epoch image pairs
flagged.  This process was iterated several times to ensure both a
clean master image, and that tips of bright stars were not incorrectly
identified as cosmic rays and removed. These master images were then used to
generate a coordinate list for the DoPHOT photometry.

Calibration of the DoPHOT photometry was carried out as follows. 
Aperture corrections were determined from the \ngc{1365} frames
and applied to the raw magnitudes.
The magnitudes were then corrected to an exposure
time of 1 second, and a zero point calibration was applied to
bring them to the 0.5 arcsec system of \markcite{holt1995a}Holtzman 
{\it et al.} (1995a).  These zero point corrections are as
given in \markcite{holt1995a}Holtzman {\it et al.}, but with a 
small correction applied to account for differences in star and sky 
apertures (\markcite{hill1998}Hill {\it et al.} 1998).
The prescription of \markcite{holt1995a}Holtzman {\it et al.}
was then used to convert the instrumental magnitudes
to standard Johnson $V$ and Cousins $I$ magnitudes. 

\subsection{Comparison of DAOPHOT and DoPHOT Photometric Systems}

The independent data reductions using ALLFRAME and DoPHOT provide a
robust external test for the accuracy of the profile-fitting photometry of
these crowded fields.  
Figure 2 shows the comparison between the DoPHOT and ALLFRAME final photometry.
As expected we see increased scatter as one goes to fainter magnitudes.
We also see that for some of the filter and chip combinations (i.e. Chip 1 $V$
and $I$ bands, and Chip 2 $V$ band) there are small scale errors, on the order
of 0.01 mag mag$^{-1}$. The nature of these scale errors is still not 
completely
clear at this time, but it is most likely due to small differences in the sky 
determination, which translate into correlations between the photometric
error and the star magnitudes. Artificial star simulations (discussed below) 
show that DoPHOT is
somewhat more robust than ALLFRAME in separating close companions, which will
therefore be measured brighter by ALLFRAME than DoPHOT. This explains the 
larger number of outlyers found with positive DoPHOT$-$ALLFRAME residuals for 
some filter/chip combinations.

The horizontal line in each panel in Figure 2 marks the average difference
between DoPHOT and ALLFRAME,
using stars brighter than 25 mag and removing wildly discrepant stars.
The differences are listed in Table 2. The errors are the rms of the means.
In general the differences are on the order of
$\pm$ 0.1 mag, which is slightly larger than DoPHOT/ALLFRAME comparisons
for other Key Project galaxies ($\sim 0.07$ mag or less for
M101, \ngc{925}, \ngc{3351}, \ngc{2090}, and \ngc{3621}). 
An extensive set of simulations was carried out adding artificial stars to the
NGC 1365 frames, with the intent of understanding the nature of the observed
differences. The main result of these observations, discussed in detail in a
subsequent paper (\markcite{lff1998b}Ferrarese {\it et al.} 1998b), 
is that the dominant cause of
uncertainty lies in the aperture corrections, while errors in sky 
determinations
and ability to resolve close companions play only a second order role. The
paucity of bright isolated stars in crowded fields makes the determination of
aperture corrections problematic at best. As an extreme example, the ALLFRAME 
aperture corrections derived for the first and second exposure of the first 
$I$ epoch differ by 0.3 mag for the PC. 
Since focus/jitter changes between consecutive orbit exposures are irrelevant, 
0.3 mag can be taken as a reasonable
estimate of the uncertainty in the aperture corrections for that particular
chip/filter/epoch combination (more typical variations in the aperture 
corrections are $\pm$ 0.07 mag for both the DoPHOT and ALLFRAME datasets). 
Similar considerations hold for the DoPHOT
aperture corrections. 
The DoPHOT$-$ALLFRAME differences observed for the bright
NGC 1365 stars are therefore found to be not significant when compared to the
uncertainty in the aperture corrections, and simply reflect the limit to which
photometric reduction can be pushed in these rather extreme fields.

We made a similar DoPHOT-ALLFRAME comparison
for the 34 Cepheids used to fit the period-luminosity relations (see Section 6).
The results of those comparisons are shown in Figure 3, and listed at the
bottom of Table 2. The DoPHOT-ALLFRAME differences for the Cepheids show more 
scatter, as expected, since they are $\sim$2 mag fainter than the brighter
nonvariable stars in Figure 2, but overall, the Cepheids mirror the nonvariable
star DoPHOT-ALLFRAME differences.

\section{Identification of Variable Stars}

Two methods were used to search for variable stars using the ALLFRAME 
dataset. In both cases, the search for variables was done using only 
the $V$ photometry. The $I$ photometry was used to help confirm variability and 
to determine $V-I$ colors. 
The first method was a search for stars with unusually high dispersion in 
their mean $V$ magnitudes. A few candidate variables were found this way. 
The more fruitful method employed a variation on the correlated variability 
test suggested by \markcite{ws1993}Welch \& Stetson (1993).  
First, the average magnitude and standard deviation 
over all epochs was calculated for each star.  Any magnitude
more than 2 standard deviations from the average was discarded. This removed 
many of the cosmic ray events. As another filter, if the magnitude 
difference between two single epoch observations was greater than 2.75 mag
the epoch was thrown out. Note that cosmic ray events that lead to
reasonably measured magnitudes (i.e. about the same as the average magnitude)
slip through these filters. 
For each pair of observations in an epoch, the difference between
each observation and the average magnitude is then calculated. The two
differences are multiplied together and summed over all epochs. 
The net result is that true variables will consistently have both
single epoch observations brighter or fainter than the average magnitude,
increasing the sum over all epochs. Random high or low observations will
tend to scatter around the average but not systematically within a single
epoch, so nonvariable stars will have lower sum values. 
Typical values of the sum were $\ge$ 1.0 for the variable candidates while
the nonvariable stars scattered around $\le$ 0.2, with increased scatter as
one went to the faintest magnitudes. Note that we discarded bad data only to
derive a variability index.

After obtaining a set of variable candidates from
the above procedure, the photometry for each candidate was plotted against 
date of observation. We were then able to note an approximate period for
the candidate as well as any observations affected by cosmic-ray events.
At the faint end, where we expected more contamination by nonvariable stars, 
we were able to exclude obvious nonvariables immediately. For candidate 
variables that passed this stage, any cosmic-ray event or otherwise
corrupted observations were removed in anticipation of determining a period 
of variation. Periods for the candidate variables were found using a
phase-dispersion minimization routine as described by
\markcite{stell1978}Stellingwerf (1978).  The resulting light curves
were checked by eye to verify the best period for each candidate.
Errors
on the periods were determined by examining changes in the light curve as 
various periods were used. When the light curve became visibly degraded
(i.e. photometry points out of phase) an upper/lower limit to the period
could safely be assigned.
These errors are subjective but provide the reader with a guide to how
well the light curves are sampled. As a final step, the local environment
of each candidate was inspected to check for severe crowding.

The search for variables in the DoPHOT reductions followed closely the
procedure described in \markcite{sh1990}Saha \& Hoessel (1990) and in
\markcite{lff1996}Ferrarese {\it et al.} (1996).  Candidates that
were classed as having a $\geq$ 99\% confidence of being variables
(based on a reduced chi-squared test) were then checked for
periodicity using a variant of the method of \markcite{lk1965}Lafler
\& Kinman (1965).  The number of spurious variables was minimized by
requiring that the reduced chi-squared statistic be greater than 2.0
when the minimum and maximum values were removed from the calculation.
The light curves of each variable candidate were then inspected
individually and any alternate minima in the phase dispersion relation
were checked to see which produced the best Cepheid light
curve. Generally the minimum in the phase dispersion plot produced the
best light curve. The image of each Cepheid candidate was also
inspected at a number of epochs. Those falling in severely crowded
regions or in areas dominated by CCD defects were also excluded.

The two lists of candidate variables, one from the ALLFRAME photometry
and one from the DoPHOT photometry, were then compared. Any candidates found
in only one dataset were located in the other dataset and checked for
variability. 
Candidates that appeared to be real variables in {\it both} the
ALLFRAME and DoPHOT datasets make up our final sample of   
52 Cepheids in \ngc{1365}. 
Finder charts for the 52 Cepheids are shown in Figures 4 and 5.
The Cepheid astrometry, periods, and period errors are given in Table 3.
The variables have been labeled V1 through V52 in order of descending period.
Column 1 in Table 3 identifies the Cepheids. Column 2 lists the CCD chip the
Cepheid is on. Chip 1 is the Planetary Camera and Chips 2-4 are the
Wide Field Camera chips. Columns 3 and 4 list the pixel position of
each Cepheid, as found on image u2s70202t (see Table 1). Columns
5 and 6 give the right ascension and declination in J2000 coordinates
for each Cepheid.  Column 7 lists the Cepheid period in days. Column 8
gives the period error in days and column 9 lists the logarithm of the
period. In addition to these 52 variables, 
there are several stars that appear to be definitely variable but for 
one reason or another did not make it into our list of definite Cepheids.
Reasons include uncertain periods, questionable environment (very crowded)
or variability seen in either the ALLFRAME or DoPHOT dataset but not both.
Positions, mean $V$ and $I$ magnitudes, and possible periods for these stars 
are given in Table A2 of the appendix.

\section{Variable Light Curves and Parameters}
  
To construct the ALLFRAME light curves, magnitudes obtained from images 
taken within a single epoch were averaged, with the resulting mean magnitude 
then plotted. Light curves for the Cepheids are shown in Figure 6 and the
ALLFRAME photometry is listed 
in Tables 4 and 5. The error bars in Figure 6 are 
averages of the two single exposure errors as reported by ALLFRAME that make up 
each epoch. In cases where one of the two single exposure magnitudes is
bad (i.e. cosmic ray event), the magnitude and error are from the surviving 
measurement. As with previous Key Project galaxies, ALLFRAME 
overestimates the error for a given photometric measurement in these 
undersampled WFPC2 images. This effect is discussed in the analysis of WFPC2 
data from M101 by \markcite{pbs1998a}Stetson {\it et al.} (1998a).
Briefly, when we
compare photometric measurements within a given epoch for \ngc{1365}
we find that the difference between them is significantly less than
the errors quoted by ALLFRAME. Typical real photometric differences
are $\pm$ 0.06 mag, at the magnitude level of the Cepheids,
for both $V$ and $I$, while ALLFRAME
reports errors significantly larger than $\pm$ 0.10 mag.

Mean $V$ magnitudes for the Cepheids were determined two different ways.
First, as in other papers in this series, since the observations were 
preselected to evenly sample a
typical 10$-$60 day Cepheid light curve 
(\markcite{free1994}Freedman {\it et al.} 1994), 
unweighted intensity averaged
mean magnitudes were calculated.  Second, phase-weighted mean intensity 
magnitudes $<m>$ were also calculated using

\begin{equation}
<m> = -2.5\log[\sum_{i}^{N} 0.5(\phi_{i+1} - 
\phi_{i-1})10^{-0.4m_{i}}]
\end{equation}

\noindent 
where $\phi$ is the phase, and the sum is over the entire light
cycle. The average difference between the unweighted and
phase-weighted intensity averaged mean $V$ magnitudes is only $-$0.027
$\pm$ 0.002 mag for the 52 Cepheids.  This difference is quite small,
as expected, since most of the Cepheids have nearly uniformly sampled
light curves.

With only four $I$ observations, total mean $I$ magnitudes were calculated
as follows. Using the $V$ and $I$ magnitudes at the four $I$ epochs,
average $V$ and average $I$ magnitudes were calculated. Then, the
difference between the four-epoch $V$ average ($<V>_{4}$) and 12-epoch
$V$ mean magnitude ($<V>_{12}$) was calculated for each Cepheid.
Since the amplitude of Cepheids in $V$ is almost exactly twice the
amplitude in $I$ (V:I = 1.00:0.51, \markcite{free1988} Freedman 1988),
the four-epoch $I$ magnitude was corrected to obtain the full
12-epoch $I$ magnitude, as follows:

\begin{equation}
<I>_{12}~=~<I>_{4} + 0.51(<V>_{12} - <V>_{4}).
\end{equation}

\noindent 
Cosmic-ray corrupted data were removed before determining mean
magnitudes.  Figure 7 shows the $I$ vs
($V-I$) color-magnitude diagram for the HST field of \ngc{1365},
highlighting the 52 Cepheids. The Cepheids fall neatly between
the well-defined blue plume and weak red plume of supergiants.

Table 6 lists derived ALLFRAME photometric parameters for the Cepheids.
Column 1 identifies the Cepheids. 
Columns 2-5 list the intensity-average and phase-weighted mean $V$
magnitudes and errors.
Columns 6-9 list the intensity-average and phase-weighted mean $I$
magnitudes and errors. 
All of the errors listed in Table 6 are mean magnitude dispersions.
They reflect the uncertainty due to the star's variability 
(the amplitude of variation)
and are not derived from the overestimated ALLFRAME errors.
Column 10 and 11 lists the intensity-averaged and phase-weighted $V-I$ 
color of each Cepheid. 
A symbol in column 12 
indicates the Cepheid was used in the Cepheid period-luminosity relation
fit to determine the distance to \ngc{1365} based on the criteria listed
in the next section. 

\section{Period-Luminosity Relations and the Distance to \ngc{1365}}

Standard period-luminosity (PL) relations for the LMC Cepheids are
adopted from \markcite{mf1991} Madore \& Freedman (1991) which
assume a true LMC distance modulus of 18.50 mag and total
line-of-sight LMC Cepheid reddening of E($B-V$) = 0.10 mag:

\begin{equation}
M_{V} = -2.76\log {\rm P} - 1.40
~~~{\rm and}~~~ M_{I} = -3.06\log {\rm P} - 1.81 .
\end{equation}

To determine the 
distance to \ngc{1365} a subset of the 52 Cepheids were chosen based on the 
following
criteria. The period of the Cepheid had to be between 10 and 47 days.
The lower limit of 10 days is common for all of the Key Project galaxies
and was chosen to avoid first overtone pulsators which have periods less 
than $\sim$ 10 days (Madore \& Freedman 1991). The longest period of 47 days 
was estimated from our observing window of 49 days and our actual sequence
of observations throughout that window. 
None of our Cepheids have periods under
10 days but 5 Cepheids have periods $> 47$ days, so they are
excluded from the fit.
Each variable also had to have a Cepheid-like light curve and
the same period, to within 10\%, in the ALLFRAME and DoPHOT datasets.
Next, to exclude Cepheids that were too crowded, each Cepheid had to 
contribute more than 50\% of
the light within a 2 pixel box surrounding it.
Our last criterion was that each Cepheid 
had to have a typical Cepheid-like color, 0.5 $\leq V-I \leq$ 1.5.
There are a total of 34 Cepheids that satisfied all of the above criteria and
they are indicated in column 11 of Table 6. These 34 Cepheids were used to
fit the PL relations and determine the distance to \ngc{1365}.

In order to avoid incompleteness bias in fitting a slope to the
\ngc{1365} data, only the zero point of the regression was fitted,
with the slope of the fit fixed to the LMC values.
The phase-weighted $V$ and $I$ PL relations are shown in Figure 8. The filled 
circles are the 34 Cepheids used to fit the PL relations, while the open 
circles are the other Cepheids. The solid line in each
figure represents the best fit to each dataset. The dashed lines drawn
at $\pm$ 0.54 mag for the $V$ PL relation and $\pm$ 0.36 mag for the
$I$ PL relation represent 2-sigma deviations from the mean PL
relations. In the absence of significant differential reddening the intrinsic
width of the Cepheid instability strip is expected to place the
\ngc{1365} Cepheids within these limits. 
We note that the full sample of 52 Cepheids are well contained within
the $V$ and $I$ instability strips in Figure 8.
From the PL fits, the apparent distance moduli to \ngc{1365}
are $\mu_{V}$ = \muv and $\mu_{I}$ = \mui,
where the errors are
calculated from the observed scatter in the \ngc{1365} PL data
themselves, appropriately reduced by the sample size of Cepheids.

The observed difference in the apparent distance moduli for \ngc{1365}
gives $\mu_{V} - \mu_{I}$ = $E(\hbox{\it V--I)}$ = \redn1365. 
The Key Project has adopted a reddening law of $R_{V} = A_{V}/E(V-I) = 2.45$
which is consistent with the work of Dean, Warren \& Cousins (1978),
\markcite{card1989}Cardelli {\it et al.} (1989) and Stanek (1996).
We therefore obtain A$_{V}$ = \av~ for this region of \ngc{1365}.
The true distance modulus to \ngc{1365} is then
\distmod, corresponding to a
linear distance of \distn1365 (internal errors only). 
To test how robust our calculated distance to \ngc{1365} was,
we also calculated the distance modulus using intensity averaged mean
magnitudes for the 34 Cepheids.
The resulting true distance modulus, 31.34 mag, is just slightly larger 
than our result using phase-weighted mean magnitudes. 
As another test we used the full compliment of 
52 Cepheids to fit the PL relations, using phase-weighted mean magnitudes,
and obtained a distance modulus of 31.35 mag. As expected we do see a
slight variation in the distance modulus depending on which mean magnitudes
are used or the sample size of Cepheids but the net result is that all of
these values are contained within our \distmod distance modulus and error.

The distance to \ngc{1365} was derived independently using the Cepheid
parameters derived from the DoPHOT reductions. 
The resulting apparent distance moduli
are $\mu_{V}$ = \domuv and  $\mu_{I}$ = \domui,
with a true modulus $\mu_{0}$ = \domod.  These differ from
the ALLFRAME moduli by $-$0.06, $-$0.05, and $-$0.05 mag,
respectively. Despite the differences in DoPHOT and ALLFRAME photometry
(Figures 2 and 3) the true distance moduli agree very well.
To better understand why this is so
we fit PL relations separately
for each WFPC2 chip, as seen in Table 7. The scatter in individual chip
apparent distance moduli is quite small for the ALLFRAME photometry and
somewhat more scattered for the DoPHOT photometry. 
Also, DoPHOT consistently measures a smaller 
reddening compared
to ALLFRAME for each chip and filter, except for Chip 2. Since over one-third
of the Cepheids used to fit the PL relations are in Chip 2, this reduces the
overall discrepancy in reddening when combining the Cepheids in all four chips.
As a test, we fit PL relations excluding the Chip 2 Cepheids. 
The resulting true
distance moduli are then 31.31 mag (no change) for ALLFRAME and 31.37 mag 
(change of +0.11 mag) for DoPHOT. The net effect of this comparison is
that the ALLFRAME and DoPHOT true distance moduli still agree within
their errors. Table 7 summarizes all of these PL fit tests. Column 1 in Table 7
indicates if the dataset used was the phase-weighted or intensity averaged
mean $V$ and $I$ magnitudes. Column 2 lists which chip subset was used (1,2,3
or 4) or all 4 chips (1-4). Column 3 gives the number of Cepheids used in
the fit. Columns 4 and 5 give the apparent distance moduli. Column 6 gives
the extinction in magnitudes. Column 7 lists the true distance moduli for
each PL fit test. The ALLFRAME results are on top, and the corresponding
DoPHOT results are at the bottom.

To check for any effects due to incompleteness at the faintest magnitudes
we split the final sample of 34 Cepheids into two sets of 17 Cepheids,
a bright and a faint sample, sorted using the $V$ phase-weighted ALLFRAME
photometry. The true distance modulus is
31.36 $\pm$ 0.11 mag for the bright sample and 31.28 $\pm$ 0.12 mag for
the faint sample. The slight changes in distance moduli seen by excluding
the brightest or faintest Cepheids are not significant compared to the
errors and we conclude that we
are not affected by incompleteness at the faintest magnitudes.

\subsection{Error Budget}

Table 8 presents the error budget for the distance to \ngc{1365}. 
The errors are classified as either random or systematic based on how we
will use 
\ngc{1365} to determine the Hubble Constant. For example, the LMC distance
modulus uncertainty and PL relation zero point uncertainties
are systematic errors because the Key Project is using the same LMC distance
modulus and Cepheid PL relation slopes for all of our
galaxies. We now discuss each source of error in Table 8 in detail.

The Key Project has adopted an LMC distance modulus of 18.50 $\pm$ 0.10 mag.
The review of published distances to the LMC, via various techniques,
by \markcite{west1997}Westerlund (1997) (his Table 2.8) indicates that the 
distance modulus
to the LMC is still uncertain at the 0.10 mag level. 
More recently, 
Gould \& Uza (1998), using observations of a light echo from
Supernova 1987A, suggest an LMC distance modulus no larger than 
18.37 $\pm$ 0.04 mag 
or 18.44 $\pm$ 0.05 mag for a circular or elliptical ring respectively. 
Panagia {\it et al.} (1998) used HST observations of the ring
around SN 1987A to obtain an LMC distance modulus of 18.58 $\pm$ 0.05 mag.
Work by Wood, Arnold, \& Sebo (1997) using models of the LMC Cepheid HV 905 
produced an LMC distance modulus of 18.51 $\pm$ 0.05 mag. Recent results from 
the MACHO Project (Alcock {\it et al.} 1997)
using double mode RR Lyrae stars give an LMC distance modulus of 
18.48 $\pm$ 0.19 mag.
In light of these recent results we feel that our adopted LMC distance modulus
and error are still valid.

\markcite{mf1991}Madore \& Freedman (1991) fit Cepheid PL
relations
to 32 LMC Cepheids using {\it BVRI} photoelectric photometry. The dispersions
for a single point about the PL relations are $\pm$ 0.27 mag for $V$ and 
$\pm$ 0.18 mag for $I$. The PL zero point errors are obtained 
by dividing by the square root of the number of Cepheids used in the fit,
giving us 
$\pm$ 0.05 mag for $V$ and $\pm$ 0.03 for $I$, items (2) and (3) in Table 8.

The LMC distance modulus and PL relation zero point uncertainties
are sources of systematic error within the Key Project, since we use
the Madore \& Freedman equations for all of the galaxies, and all galaxy
determinations will change systematically as improvements to the zero point
become available. 
Combining items (1), (2), and (3) in Table 8 in quadrature we obtain a PL 
relation uncertainty of $\pm$ 0.12 mag.

\markcite{hill1998}Hill {\it et al.} (1998) estimated our WFPC2 zero point 
uncertainties to be 
$\pm$ 0.02 mag for both $V$ and $I$ by comparing ground-based and HST 
observations of M100. In addition we must include the uncertainties in the 
aperture corrections.  
While there is one case where two aperture corrections within an epoch
differed by 0.3 mag (Chip 1 $I$ band, first epoch), the scatter about the mean 
ALLFRAME aperture corrections is 
$\pm$ 0.07 mag for $V$ and $\pm$ 0.06 mag for $I$.
We will take these to be the errors in zero point due to the aperture
corrections. We then combine the \markcite{hill1998}Hill {\it et al.} errors 
and the aperture
correction errors in quadrature to obtain WFPC2 zero point errors of 
$\pm$ 0.07 mag in $V$ and $\pm$ 0.06 mag in $I$, items (4) and (5) in Table 8. 
Since the photometric errors in the two bands are uncorrelated
we combine items (4) and (5) in quadrature,
but we must weight these errors, $\sigma_{V}$ and $\sigma_{I}$, by the 
differing effects of reddening, as given by
$[(1-R)^{2}(\sigma_{V})^{2} + R^{2}(\sigma_{I})^{2}]^{1/2}$.
As stated previously, we have adopted a reddening law of
$R_{V}= A_{V}/E(V-I)$ = 2.45. Our photometric contribution to the error in
the distance modulus is therefore $\pm$ 0.18 mag.

Early work by the Key Project discovered what appeared to be a long versus
short exposure zero point offset, such that stars in exposures 
longer than approximately
1000 seconds were systematically brighter, by approximately 0.05 mag, than
stars in exposures of less than 1000 seconds duration. The effect is now
thought to be a charge transfer efficiency effect in the WFPC2 CCDs. This
effect is discussed in detail in \markcite{hill1998}Hill {\it et al.} (1998) 
and also in
\markcite{wh1997}Whitmore \& Heyer (1997). For the Key Project
we have used the long exposure zero point for our photometric calibration.
At this time, an offset of +0.05 mag for both $V$ and $I$ is thought to be the 
best estimate to correct for this effect. 
Additional work on the zero points is currently
being undertaken by \markcite{pbs1998b}Stetson {\it et al.} (1998b) 
based on an 
extensive set of ground based and HST data.
This effect is included in Table 8 as a source of systematic error in the
distance to \ngc{1365}.

There is an additional distance modulus error introduced from fitting the
Cepheid PL relations. 
From Section 5, these errors are $\pm$ 0.05 mag
in $V$ and $\pm$ 0.06 mag in $I$, and  include photometric errors,
differential reddening and the apparent width of the instability strip in
both $V$ and $I$. These errors are combined in quadrature in item (R2)
in Table 8.

There is concern that the Cepheid PL relation may have a metallicity
dependence. The Cepheids in the calibrating LMC are thought to be
relatively metal poor compared Fornax and Virgo cluster galaxies
(Kennicutt {\it et al.} 1998),
so a metallicity dependence in the Cepheid PL relation
would be a source of systematic error in our distance determination.
Kennicutt {\it et al.} measured
a marginal metallicity dependence in one of the Key Project
galaxies, M101, leading to a
shift in distance modulus of 
$\delta(m-M)_0/\delta[O/H] = -0.24 \pm 0.16$ mag/dex.
We can use this relation to estimate the systematic error in distance modulus 
to \ngc{1365} due to a metallicity dependence on the Cepheid PL relations.
Zaritsky, Kennicutt \& Huchra (1994)
found the oxygen abundance in \ngc{1365} to be $12 + {\rm log}(O/H) = 9.0$
while the measured oxygen
abundance in the LMC on this scale is $12 + {\rm log}(O/H) = 8.50$ 
(Kennicutt {\it et al.} 1998).
The 0.5 dex difference in oxygen abundance would, if applied, 
increase the true distance modulus by +0.12 mag or a 6\% 
increase in the distance to \ngc{1365}.

The total uncertainty in distance modulus to \ngc{1365} due to random
errors is obtained by combining the distance modulus and PL fit uncertainties
in quadrature:
$[0.18^{2} + 0.08^{2}]^{1/2}$ = $\pm$ 0.20 mag.

The total systematic error in the distance to \ngc{1365} is obtained by
combining the systematic errors, the LMC Cepheid PL relation, a possible 
metallicity dependence, and the long versus short exposure zero point,
in quadrature:
$[0.12^{2} + 0.12^{2}+0.05^{2}]^{1/2}$ = $\pm$ 0.18 mag.
Thus our final distance modulus to \ngc{1365} is 
31.31 $\pm$ 0.20 (random) $\pm$ 0.18 (systematic).
Th implications of a Cepheid distance to \ngc{1365} and the Fornax
cluster are discussed in the companion paper Madore {\it et al.} (1998).

\section{Conclusion}
We have used the HST WFPC2 instrument to obtain 12 epochs in $V$ and
4 epochs in $I$ of the eastern part of \ngc{1365} located in the Fornax Cluster.
The images were reduced separately using the ALLFRAME and DoPHOT 
photometry packages. The raw photometry was transformed to standard
Johnson $V$ and Cousins $I$ photometry via Hill {\it et al.} (1998)
zero points, Holtzman {\it et al.} (1995a) color terms, and 
aperture corrections
derived from the \ngc{1365} images. The $\lesssim$ 0.1 mag difference
between the final ALLFRAME and DoPHOT photometry is most likely due to
the relatively large uncertainties in both the ALLFRAME and DoPHOT
aperture corrections, which show a scatter of $\pm$ 0.07 mag. 

A total of 52 Cepheids
were discovered in \ngc{1365}, ranging in period from 14 to 60 days. 
A subset of 34 Cepheids was chosen based on period, light curve
shape, color and relative crowding to fit Cepheid period-luminosity relations
using fixed slopes from 
\markcite{mf1991} Madore \& Freedman (1991). A distance modulus of
31.31 $\pm$ 0.20 (random) $\pm$ 0.18 (systematic) mag, corresponding to 
18.3 $\pm$ 1.7 (random) $\pm$ 1.6 (systematic) Mpc is obtained from the 
ALLFRAME photometry. A distance modulus of \domod (internal errors only) 
was obtained from the DoPHOT photometry.
The average reddening in this region of \ngc{1365} was measured to be
$E(\hbox{\it V--I)}$ = \redn1365 using ALLFRAME and 0.15 $\pm$ 0.10 mag 
using DoPHOT.
The apparent agreement in distance between ALLFRAME and DoPHOT despite the
observed $\lesssim$ 0.1 mag photometric differences is due to DoPHOT measuring
a lower reddening to \ngc{1365} in WFPC2 Chips 1, 3 and 4 and a higher
reddening than ALLFRAME in Chip 2. Over one-third of the Cepheids in the
PL fit sample are found in Chip 2, reducing the {\it mean} difference in
reddening and distance modulus between DoPHOT and ALLFRAME.
The companion paper by Madore {\it et al.} (1998) discusses in detail
the implications of a Cepheid distance to \ngc{1365}.

\acknowledgments
The work presented in this paper is based on observations made by the
NASA/ESA Hubble Space Telescope, obtained by the Space Telescope
Science Institute, which is operated by AURA, Inc. under NASA contract
No. 5-26555.  We gratefully acknowledge the support of the NASA and
STScI support staff, with special thanks our program coordinator, Doug
Van Orsow.  Support for this work was provided by NASA through grant
GO-2227-87A from STScI. 
This paper is based partially on data obtained at the Las Campanas
Observatory 2.5m telescope in Chile, owned and operated by the Carnegie
Institution of Washington.
SMGH and PBS are grateful to NATO for travel support
via a Collaborative Research Grant (960178).
LF acknowledges support by NASA through Hubble Fellowship grant
HF-01081.01-96A awarded by the Space Telescope Science Institute,
which is operated by AURA, Inc., for NASA under contract NSA 5-26555.
The research described in this paper was partially carried out by the Jet
Propulsion Laboratory, California Institute of Technology, under a
contract with the National Aeronautics and Space Administration.  This
research has made use of the NASA/IPAC Extragalactic Database (NED)
which is operated by the Jet Propulsion Laboratory, California
Institute of Technology, under a contract with the National
Aeronautics and Space Administration. 

\appendix

\section{Appendix}

\subsection{Secondary Standard Photometry}
Several bright stars in each WFPC2 field were used to determine the
ALLFRAME aperture corrections. These relatively isolated stars are 
presented here as secondary standards for our \ngc{1365} HST field.

\subsection{Possible Variable Stars}
There are several stars within the ALLFRAME and DoPHOT datasets that appear 
to be variable but for 
one reason or another did not make it into our list of definite Cepheids.
Reasons may include uncertain periods, too crowded, extremely red or blue,
or the variability is seen in either the ALLFRAME or DoPHOT dataset but not 
both. Positions, ALLFRAME mean $V$ and $I$ intensity averaged magnitudes, 
and possible periods for these stars are given in the table below.

\newpage

\begin{deluxetable}{ccccc}
\tablenum{1}
\tablewidth{27pc}
\tablecaption{HST Observations of NGC 1365}
\tablehead{
\colhead{Image}&
\colhead{UT Day}& 
\colhead{HJD}&
\colhead{Exp. Time}&
\colhead{HST}\\
\colhead{Identification}&
\colhead{1995}&
\colhead{2449000.+}&
\colhead{(seconds)} &\colhead{Filter}
}
\startdata 
u2s70201t & Aug 6 &   936.1312 & 2400 & F555W \nl
u2s70202t & Aug 6 &   936.1961 & 2700 & F555W \nl
u2s70203t & Aug 6 &   936.2635 & 2700 & F814W \nl
u2s70204t & Aug 6 &   936.3301 & 2700 & F814W \nl
u2s70301t & Aug 14 &  943.9046 & 2400 & F555W \nl
u2s70302t & Aug 14 &  943.9688 & 2700 & F555W \nl
u2s70303t & Aug 15 &  944.0355 & 2700 & F814W \nl
u2s70304t & Aug 15 &  944.1028 & 2700 & F814W \nl
u2s70401t & Aug 21 &  951.2786 & 2400 & F555W \nl
u2s70402t & Aug 21 &  951.3407 & 2700 & F555W \nl
u2s70501t & Aug 24 &  953.6266 & 2400 & F555W \nl
u2s70502t & Aug 24 &  953.6915 & 2700 & F555W \nl
u2s70601t & Aug 26 &  956.1726 & 2400 & F555W \nl
u2s70602t & Aug 26 &  956.2362 & 2700 & F555W \nl
u2s70701t & Aug 29 &  959.1880 & 2400 & F555W \nl
u2s70703t & Aug 29 &  959.2583 & 2300 & F555W \nl
u2s70705t & Aug 29 &  959.3256 & 2300 & F814W \nl
u2s70706t & Aug 29 &  959.3897 & 2700 & F814W \nl
u2s70801t & Sept 1 &  962.2771 & 2400 & F555W \nl
u2s70802t & Sept 1 &  962.3413 & 2700 & F555W \nl
u2s70901t & Sept 5 &  966.6966 & 2400 & F555W \nl
u2s70902t & Sept 5 &  966.7602 & 2700 & F555W \nl
u2s71001t & Sept 8 &  969.3816 & 2400 & F555W \nl
u2s71002t & Sept 8 &  969.4465 & 2700 & F555W \nl
u2s71101t & Sept 13 & 974.2040 & 2400 & F555W \nl
u2s71102t & Sept 13 & 974.2717 & 2700 & F555W \nl
u2s71201t & Sept 19 & 980.2369 & 2400 & F555W \nl
u2s71202t & Sept 19 & 980.3026 & 2700 & F555W \nl
u2s71203t & Sept 19 & 980.3692 & 2700 & F814W \nl
u2s71204t & Sept 19 & 980.4359 & 2700 & F814W \nl
u2s71301t & Sept 24 & 985.2620 & 2400 & F555W \nl
u2s71302t & Sept 24 & 985.3277 & 2700 & F555W \nl
\enddata
\end{deluxetable}

\begin{deluxetable}{ccccc}
\tablenum{2}
\tablewidth{40pc}
\tablecaption{DoPHOT vs ALLFRAME Photometry}
\tablehead{
\colhead{ } &
\colhead{Number} &
\colhead{DoPHOT-ALLFRAME}&
\colhead{Number} &
\colhead{DoPHOT-ALLFRAME}\\
\colhead{Chip}&
\colhead{of Stars}&
\colhead{$\Delta$V} &
\colhead{of Stars}&
\colhead{$\Delta$I}
}
\startdata
\multicolumn{5}{c}{bright non-variable stars} \nl
1 & 103 &   +0.13 $\pm$ 0.04 & 329 &   +0.10 $\pm$ 0.06\nl
2 & 332 &   +0.02 $\pm$ 0.06 & 649 & $-$0.11 $\pm$ 0.05\nl
3 &  40 & $-$0.09 $\pm$ 0.03 & 179 & $-$0.15 $\pm$ 0.04\nl
4 & 159 &   +0.03 $\pm$ 0.07 & 401 & $-$0.07 $\pm$ 0.04\nl
&&&&\nl
\multicolumn{5}{c}{Cepheids} \nl
1& 11 & $-$0.01 $\pm$ 0.04 &11& +0.04 $\pm$ 0.04 \nl
2& 12 & $-$0.14 $\pm$ 0.02 &12& $-$0.18 $\pm$ 0.02 \nl
3& 8  & $-$0.11 $\pm$ 0.02 & 8& $-$0.11 $\pm$ 0.05 \nl
4& 6  & $-$0.08 $\pm$ 0.05 & 6& $-$0.05 $\pm$ 0.06 \nl
\enddata
\end{deluxetable}

\tablenum{3}
\tablewidth{35pc}
\begin{deluxetable}{ccccccccc}
\tablecaption{Positions and Periods of NGC 1365 Cepheids}
\tablehead{
\colhead{ID}& \colhead{Chip} & \colhead{x} & \colhead{y} & 
\colhead{RA(J2000)}& \colhead{Dec(J2000)} &
\colhead{P(days)} & \colhead{$\sigma$ P} & \colhead{logP}}
\startdata
V1  & 1 & 501.7 & 245.8 & 3:33:43.72 & -36:09:21.6 &60.0& \nodata & 1.78  \nl
V2  & 4 & 480.7 & 446.9 & 3:33:45.28 & -36:10:13.6 &60.0& \nodata & 1.78  \nl
V3  & 3 & 451.6 & 481.6 & 3:33:50.44 & -36:09:20.7 &55.0& \nodata & 1.74  \nl
V4  & 1 & 769.4 & 428.3 & 3:33:42.51 & -36:09:23.2 &55.0& \nodata & 1.74  \nl
V5  & 1 & 294.4 & 183.4 & 3:33:44.47 & -36:09:17.7 &53.0& \nodata & 1.72  \nl
V6  & 1 & 590.2 & 423.4 & 3:33:43.04 & -36:09:18.1 &  47.0 & 4.0 & 1.67   \nl
V7  & 4 & 415.3 & 666.8 & 3:33:43.55 & -36:10:22.7 &  44.2 & 5.2 & 1.65   \nl
V8  & 4 & 479.5 & 486.8 & 3:33:45.03 & -36:10:16.1 &  42.5 & 4.5 & 1.63   \nl
V9  & 1 & 428.2 & 228.2 & 3:33:43.98 & -36:09:20.1 &  41.5 & 3.5 & 1.62   \nl
V10 & 3 & 513.3 & 457.7 & 3:33:50.70 & -36:09:14.9 &  41.4 & 3.6 & 1.62   \nl
V11 & 2 & 424.5 & 287.5 & 3:33:45.13 & -36:08:29.7 &  41.0 & 7.0 & 1.61   \nl
V12 & 2 & 400.9 & 617.7 & 3:33:47.31 & -36:08:10.0 &  39.2 & 4.8 & 1.59   \nl
V13 & 2 &  89.5 & 378.2 & 3:33:47.49 & -36:08:49.1 &  38.0 & 4.0 & 1.58   \nl
V14 & 3 & 200.5 & 285.7 & 3:33:47.82 & -36:09:22.1 &  37.6 & 2.6 & 1.58   \nl
V15 & 4 & 586.4 & 695.2 & 3:33:44.27 & -36:10:37.6 &  35.2 & 1.2 & 1.55   \nl
V16 & 1 & 456.2 & 724.5 & 3:33:42.69 & -36:09:03.7 &  35.1 & 3.1 & 1.55   \nl
V17 & 2 & 244.1 & 571.1 & 3:33:47.86 & -36:08:24.9 &  35.0 & 1.5 & 1.54   \nl
V18 & 3 & 130.3 & 732.3 & 3:33:49.76 & -36:10:00.4 &  34.6 & 1.9 & 1.54   \nl
V19 & 1 & 162.7 & 432.9 & 3:33:44.24 & -36:09:05.1 &  34.5 & 4.5 & 1.54   \nl
V20 & 2 & 423.6 & 663.7 & 3:33:47.47 & -36:08:05.3 &  34.4 & 4.4 & 1.54   \nl
V21 & 3 & 311.8 & 612.5 & 3:33:50.26 & -36:09:39.6 &  34.0 & 2.0 & 1.53   \nl
V22 & 3 & 120.6 & 226.0 & 3:33:47.00 & -36:09:22.7 &  34.0 & 2.5 & 1.53   \nl
V23 & 1 & 691.8 & 112.0 & 3:33:43.50 & -36:09:31.9 &  34.0 & 2.3 & 1.53   \nl
V24 & 4 & 500.6 & 464.7 & 3:33:45.28 & -36:10:16.3 &  33.6 & 0.6 & 1.53   \nl
V25 & 1 & 292.1 & 430.9 & 3:33:43.87 & -36:09:09.0 &  33.1 & 2.5 & 1.52   \nl
V26 & 2 & 306.0 & 619.8 & 3:33:47.83 & -36:08:17.0 &  32.5 & 2.5 & 1.51   \nl
V27 & 2 & 424.3 & 574.6 & 3:33:46.91 & -36:08:11.1 &  31.3 & 3.3 & 1.50   \nl
V28 & 3 & 449.9 & 396.8 & 3:33:49.97 & -36:09:14.4 &  30.3 & 4.0 & 1.48   \nl
V29 & 4 & 156.9 & 595.9 & 3:33:42.63 & -36:09:58.4 &  29.8 & 4.2 & 1.47   \nl
V30 & 2 & 438.1 & 139.8 & 3:33:44.13 & -36:08:38.3 &  29.6 & 2.7 & 1.47   \nl
V31 & 4 & 390.7 & 359.9 & 3:33:45.36 & -36:10:01.2 &  29.0 & 2.3 & 1.46   \nl
V32 & 3 & 392.3 & 339.6 & 3:33:49.31 & -36:09:13.8 &  28.9 & 2.2 & 1.46   \nl
V33 & 2 & 517.7 &  39.1 & 3:33:43.08 & -36:08:38.9 &  28.0 & 2.0 & 1.45   \nl
V34 & 1 & 706.4 & 473.5 & 3:33:42.58 & -36:09:19.7 &  28.0 & 3.0 & 1.45   \nl
V35 & 4 & 295.8 & 123.9 & 3:33:46.35 & -36:09:38.8 &  27.5 & 4.0 & 1.44   \nl
V36 & 2 & 395.2 & 170.1 & 3:33:44.55 & -36:08:39.6 &  26.6 & 2.6 & 1.42   \nl
V37 & 4 & 168.9 & 545.3 & 3:33:43.02 & -36:09:56.1 &  26.5 & 1.5 & 1.42   \nl
V38 & 3 & 630.0 & 499.6 & 3:33:51.65 & -36:09:10.5 &  26.5 & 1.5 & 1.42   \nl
V39 & 2 & 170.5 & 478.3 & 3:33:47.68 & -36:08:36.4 &  26.1 & 3.9 & 1.42   \nl
V40 & 1 & 774.3 & 292.5 & 3:33:42.83 & -36:09:28.0 &  25.0 & 1.0 & 1.40   \nl
V41 & 1 & 744.9 & 158.7 & 3:33:43.24 & -36:09:31.8 &  23.5 & 1.0 & 1.37   \nl
V42 & 2 &  82.7 & 366.7 & 3:33:47.45 & -36:08:50.3 &  22.9 & 1.9 & 1.36   \nl
V43 & 2 & 417.5 & 622.9 & 3:33:47.25 & -36:08:08.4 &  21.0 & 1.0 & 1.32   \nl
V44 & 2 & 604.4 & 610.8 & 3:33:46.17 & -36:07:55.1 &  21.0 & 0.7 & 1.32   \nl
V45 & 2 & 542.6 & 690.6 & 3:33:47.00 & -36:07:54.6 &  20.3 & 1.4 & 1.31   \nl
V46 & 1 & 247.1 & 124.1 & 3:33:44.75 & -36:09:18.3 &  20.1 & 1.6 & 1.30   \nl
V47 & 3 & 238.4 & 609.4 & 3:33:49.79 & -36:09:44.1 &  20.0 & 1.0 & 1.30   \nl
V48 & 4 & 159.0 &  93.6 & 3:33:45.82 & -36:09:26.4 &  18.8 & 1.1 & 1.27   \nl
V49 & 1 & 475.3 & 217.9 & 3:33:43.86 & -36:09:21.8 &  18.0 & 0.6 & 1.26   \nl
V50 & 4 & 233.5 & 632.1 & 3:33:42.81 & -36:10:06.6 &  16.3 & 1.7 & 1.21   \nl
V51 & 3 & 256.1 & 465.1 & 3:33:49.13 & -36:09:32.1 &  16.2 & 0.3 & 1.21   \nl
V52 & 3 & 322.7 & 187.8 & 3:33:48.07 & -36:09:06.8 &  14.2 & 0.4 & 1.15   \nl
\enddata
\end{deluxetable}

\tablenum{4}
\tablewidth{42pc}
\begin{deluxetable}{ccccccc}
\tablecaption{V Photometry for NGC 1365 Cepheids}
\tablehead{
\colhead{}&
\multicolumn{6}{c}{Heliocentric Julian Date 2449000.+}\nl
\colhead{Cepheid}&
\colhead{936.164}&
\colhead{943.937}&
\colhead{951.310}&
\colhead{953.659}&
\colhead{956.204}&
\colhead{959.223}}
\startdata
   V1& 25.29 $\pm$  0.12& 25.21 $\pm$  0.13& 25.47 $\pm$  0.19& 25.50 $\pm$  0.14& 25.44 $\pm$  0.17& 25.49 $\pm$  0.16\nl
   V2& 25.40 $\pm$  0.19& 25.22 $\pm$  0.13& 24.76 $\pm$  0.15& 24.63 $\pm$  0.14& 24.61 $\pm$  0.14& 24.62 $\pm$  0.13\nl
   V3& 25.44 $\pm$  0.13& 25.20 $\pm$  0.15& 24.38 $\pm$  0.19& 24.53 $\pm$  0.14& 24.50 $\pm$  0.11& 24.43 $\pm$  0.12\nl
   V4& 25.31 $\pm$  0.15& 25.37 $\pm$  0.16& 25.66 $\pm$  0.17& 25.62 $\pm$  0.26& 25.63 $\pm$  0.20& 25.69 $\pm$  0.16\nl
   V5& 25.86 $\pm$  0.15& 25.97 $\pm$  0.17& 26.29 $\pm$  0.21& 26.29 $\pm$  0.16& 26.12 $\pm$  0.20& 26.33 $\pm$  0.24\nl
   V6& 26.44 $\pm$  0.26& 25.90 $\pm$  0.26& 25.68 $\pm$  0.16& 25.57 $\pm$  0.14& 25.57 $\pm$  0.21& 25.94 $\pm$  0.20\nl
   V7& 26.23 $\pm$  0.29& 25.02 $\pm$  0.12& 25.32 $\pm$  0.17& 25.47 $\pm$  0.13& 25.57 $\pm$  0.16& 25.54 $\pm$  0.13\nl
   V8& 26.02 $\pm$  0.15& 25.04 $\pm$  0.13& 25.35 $\pm$  0.18& 25.44 $\pm$  0.16& 25.58 $\pm$  0.16& 25.50 $\pm$  0.15\nl
   V9& 25.95 $\pm$  0.18& 26.27 $\pm$  0.19& 26.21 $\pm$  0.19&  \nodata         & 25.21 $\pm$  0.45& 25.50 $\pm$  0.14\nl
  V10& 25.72 $\pm$  0.15& 25.56 $\pm$  0.29& 26.00 $\pm$  0.20&  \nodata         & 25.97 $\pm$  0.15& 26.03 $\pm$  0.19\nl
  V11& 26.65 $\pm$  0.35& 27.01 $\pm$  0.48& 27.16 $\pm$  0.63& 26.12 $\pm$  0.26& 26.02 $\pm$  0.25& 25.97 $\pm$  0.21\nl
  V12& 26.18 $\pm$  0.21& 25.52 $\pm$  0.15& 25.80 $\pm$  0.19& 25.97 $\pm$  0.18& 25.99 $\pm$  0.17& 25.79 $\pm$  0.13\nl
  V13& 25.84 $\pm$  0.18& 26.17 $\pm$  0.20& 26.62 $\pm$  0.31& 26.52 $\pm$  0.29& 26.74 $\pm$  0.32& 26.73 $\pm$  0.29\nl
  V14& 26.73 $\pm$  0.22& 26.85 $\pm$  0.30& 26.02 $\pm$  0.17& 26.25 $\pm$  0.20& 26.17 $\pm$  0.17& 26.06 $\pm$  0.52\nl
  V15& 26.12 $\pm$  0.17& 26.34 $\pm$  0.21& 25.62 $\pm$  0.19& 25.66 $\pm$  0.20& 25.59 $\pm$  0.16& 25.68 $\pm$  0.16\nl
  V16& 25.78 $\pm$  0.19& 25.86 $\pm$  0.16& 26.07 $\pm$  0.24& 26.32 $\pm$  0.22& 26.32 $\pm$  0.23& 26.15 $\pm$  0.20\nl
  V17& 25.97 $\pm$  0.15& 26.15 $\pm$  0.24& 25.12 $\pm$  0.16& 25.21 $\pm$  0.27& 25.38 $\pm$  0.18& 25.35 $\pm$  0.14\nl
  V18& 25.53 $\pm$  0.13& 25.80 $\pm$  0.18& 26.24 $\pm$  0.40& 26.34 $\pm$  0.25& 26.16 $\pm$  0.17& 25.70 $\pm$  0.14\nl
  V19& 25.97 $\pm$  0.19& 26.47 $\pm$  0.21& 26.72 $\pm$  0.30& 26.57 $\pm$  0.25& 26.42 $\pm$  0.23& 25.66 $\pm$  0.18\nl
  V20& 26.38 $\pm$  0.16& 26.84 $\pm$  0.26& 25.77 $\pm$  0.15& 25.48 $\pm$  0.13& 25.65 $\pm$  0.16& 25.84 $\pm$  0.16\nl
  V21& 25.76 $\pm$  0.16& 25.89 $\pm$  0.23& 26.30 $\pm$  0.22& 26.53 $\pm$  0.21& 26.52 $\pm$  0.20& 26.24 $\pm$  0.16\nl
  V22& 26.29 $\pm$  0.41& 25.66 $\pm$  0.15& 25.77 $\pm$  0.15& 25.81 $\pm$  0.19& 25.79 $\pm$  0.17& 25.88 $\pm$  0.15\nl
  V23& 26.19 $\pm$  0.17& 25.29 $\pm$  0.13& 25.68 $\pm$  0.17& 26.02 $\pm$  0.23& 25.76 $\pm$  0.15& 25.89 $\pm$  0.19\nl
  V24& 26.22 $\pm$  0.19& 26.74 $\pm$  0.30&  \nodata         & 25.82 $\pm$  0.16& 25.64 $\pm$  0.46& 25.59 $\pm$  0.22\nl
  V25& 26.06 $\pm$  0.18& 26.58 $\pm$  0.22& 25.31 $\pm$  0.13& 25.55 $\pm$  0.16& 25.52 $\pm$  0.16& 25.81 $\pm$  0.15\nl
  V26& 25.53 $\pm$  0.17& 25.82 $\pm$  0.19& 26.23 $\pm$  0.24& 26.40 $\pm$  0.25& 26.33 $\pm$  0.31& 25.51 $\pm$  0.16\nl
  V27& 26.61 $\pm$  0.24& 25.49 $\pm$  0.13& 26.10 $\pm$  0.23& 26.26 $\pm$  0.19& 26.34 $\pm$  0.19& 26.33 $\pm$  0.26\nl
  V28& 26.67 $\pm$  0.23& 26.64 $\pm$  0.23& 26.43 $\pm$  0.26& 26.21 $\pm$  0.18& 26.41 $\pm$  0.15& 26.36 $\pm$  0.19\nl
  V29& 26.32 $\pm$  0.26& 26.45 $\pm$  0.21& 25.60 $\pm$  0.19& 25.67 $\pm$  0.18& 25.72 $\pm$  0.16& 25.78 $\pm$  0.18\nl
  V30& 25.56 $\pm$  0.16& 25.84 $\pm$  0.21& 25.98 $\pm$  0.23& 25.55 $\pm$  0.23& 25.05 $\pm$  0.13& 24.87 $\pm$  0.35\nl
  V31& 25.73 $\pm$  0.32& 26.56 $\pm$  0.19& 26.70 $\pm$  0.27& 26.09 $\pm$  0.24& 25.61 $\pm$  0.15& 25.72 $\pm$  0.15\nl
  V32&  \nodata         & 26.58 $\pm$  0.23& 26.35 $\pm$  0.22& 26.81 $\pm$  0.28& 26.75 $\pm$  0.24& 26.77 $\pm$  0.28\nl
  V33& 25.77 $\pm$  0.22& 25.79 $\pm$  0.29& 25.65 $\pm$  0.63& 25.79 $\pm$  0.33& 25.27 $\pm$  0.23& 25.40 $\pm$  0.19\nl
  V34& 25.81 $\pm$  0.55& 26.10 $\pm$  0.25& 25.63 $\pm$  0.16& 25.65 $\pm$  0.13& 25.71 $\pm$  0.16& 25.74 $\pm$  0.32\nl
  V35& 26.80 $\pm$  0.26& 27.11 $\pm$  0.25& 25.99 $\pm$  0.27& 26.21 $\pm$  0.19& 26.16 $\pm$  0.18& 26.44 $\pm$  0.23\nl
  V36& 26.77 $\pm$  0.41& 26.00 $\pm$  0.22& 26.73 $\pm$  0.37& 26.58 $\pm$  0.52& 26.47 $\pm$  0.25& 26.94 $\pm$  0.33\nl
  V37& 25.69 $\pm$  0.15& 26.24 $\pm$  0.22& 27.12 $\pm$  0.45& 26.88 $\pm$  0.31& 26.47 $\pm$  0.27& 25.72 $\pm$  0.31\nl
  V38& 26.79 $\pm$  0.25& 25.95 $\pm$  0.18& 26.49 $\pm$  0.19& 26.50 $\pm$  0.18& 26.49 $\pm$  0.35& 26.53 $\pm$  0.22\nl
  V39& 26.44 $\pm$  0.24& 25.62 $\pm$  0.15& 26.17 $\pm$  0.22& 26.39 $\pm$  0.19& 26.58 $\pm$  0.23& 26.39 $\pm$  0.21\nl
  V40& 25.46 $\pm$  0.31& 25.88 $\pm$  0.26& 26.24 $\pm$  0.35& 26.26 $\pm$  0.35& 26.32 $\pm$  0.44& 26.30 $\pm$  0.33\nl
  V41& 26.05 $\pm$  0.22& 25.36 $\pm$  0.15& 26.07 $\pm$  0.30& 26.15 $\pm$  0.28& 26.16 $\pm$  0.23& 26.14 $\pm$  0.25\nl
  V42& 26.37 $\pm$  0.19& 25.59 $\pm$  0.14& 26.28 $\pm$  0.19& 26.25 $\pm$  0.21& 26.34 $\pm$  0.27& 26.47 $\pm$  0.25\nl
  V43& 26.48 $\pm$  0.24& 26.13 $\pm$  0.20& 26.58 $\pm$  0.27& 26.62 $\pm$  0.25& 26.10 $\pm$  0.46& 25.62 $\pm$  0.47\nl
  V44& 26.95 $\pm$  0.26& 26.22 $\pm$  0.20& 27.53 $\pm$  0.52& 27.18 $\pm$  0.38& 27.08 $\pm$  0.37& 25.73 $\pm$  0.14\nl
  V45& 27.36 $\pm$  0.57& 26.48 $\pm$  0.19& 27.06 $\pm$  0.37& 27.16 $\pm$  0.36& 27.33 $\pm$  0.45& 27.11 $\pm$  0.38\nl
  V46& 26.75 $\pm$  0.23& 25.65 $\pm$  0.14& 26.32 $\pm$  0.18& 26.63 $\pm$  0.21& 26.53 $\pm$  0.50& 26.61 $\pm$  0.70\nl
  V47& 25.93 $\pm$  0.15& 26.42 $\pm$  0.18& 26.45 $\pm$  0.36& 25.77 $\pm$  0.17& 26.04 $\pm$  0.18& 25.81 $\pm$  0.44\nl
  V48& 26.08 $\pm$  0.17& 26.89 $\pm$  0.26& 27.11 $\pm$  0.31& 25.94 $\pm$  0.18& 26.24 $\pm$  0.14& 26.46 $\pm$  0.25\nl
  V49& 26.63 $\pm$  0.25& 27.17 $\pm$  0.40& 26.44 $\pm$  0.23& 26.50 $\pm$  0.25& 26.96 $\pm$  0.29& 27.20 $\pm$  0.33\nl
  V50& 26.66 $\pm$  0.28& 26.37 $\pm$  0.24& 26.69 $\pm$  0.29& 26.80 $\pm$  0.26& 27.07 $\pm$  0.37& 26.93 $\pm$  0.35\nl
  V51& 26.97 $\pm$  0.28& 26.86 $\pm$  0.47& 27.45 $\pm$  0.47& 26.49 $\pm$  0.21& 26.79 $\pm$  0.36& 26.74 $\pm$  0.21\nl
  V52& 27.36 $\pm$  0.34& 26.55 $\pm$  0.21& 27.12 $\pm$  0.32& 27.59 $\pm$  0.47& 26.31 $\pm$  0.21& 26.42 $\pm$  0.21\nl
\enddata
\end{deluxetable}

\tablenum{4}
\tablewidth{42pc}
\begin{deluxetable}{ccccccc}
\tablecaption{V Photometry for NGC 1365 Cepheids}
\tablehead{
\colhead{}&
\multicolumn{6}{c}{Heliocentric Julian Date 2449000.+}\nl
\colhead{Cepheid}&
\colhead{962.309}&
\colhead{965.728}&
\colhead{969.414}&
\colhead{974.238}&
\colhead{980.270}&
\colhead{985.295}}
\startdata
   V1& 25.48 $\pm$  0.13& 25.66 $\pm$  0.13& 25.65 $\pm$  0.13& 25.90 $\pm$  0.14& 25.94 $\pm$  0.18& 26.06 $\pm$  0.19\nl
   V2& 24.61 $\pm$  0.12& 24.75 $\pm$  0.11& 24.76 $\pm$  0.12& 24.80 $\pm$  0.13& 24.99 $\pm$  0.11& 25.00 $\pm$  0.14\nl
   V3& 24.65 $\pm$  0.12& 24.80 $\pm$  0.11& 24.75 $\pm$  0.22& 24.81 $\pm$  0.11& 24.96 $\pm$  0.12& 25.16 $\pm$  0.13\nl
   V4& 25.80 $\pm$  0.16& 25.71 $\pm$  0.15& 25.73 $\pm$  0.17& 25.44 $\pm$  0.13& 25.17 $\pm$  0.14& 25.20 $\pm$  0.14\nl
   V5& 26.35 $\pm$  0.45& 26.47 $\pm$  0.31& 26.68 $\pm$  0.17& 26.48 $\pm$  0.21& 25.53 $\pm$  0.31& 25.79 $\pm$  0.17\nl
   V6& 25.86 $\pm$  0.21& 25.98 $\pm$  0.18& 26.11 $\pm$  0.19& 26.19 $\pm$  0.18& 26.61 $\pm$  0.26& 26.16 $\pm$  0.20\nl
   V7& 25.61 $\pm$  0.35& 25.53 $\pm$  0.42& 25.85 $\pm$  0.32& 26.14 $\pm$  0.17& 26.06 $\pm$  0.16& 24.70 $\pm$  0.17\nl
   V8& 25.62 $\pm$  0.31& 25.97 $\pm$  0.18& 25.88 $\pm$  0.18& 25.99 $\pm$  0.20& 26.09 $\pm$  0.20& 24.94 $\pm$  0.14\nl
   V9& 25.43 $\pm$  0.16& 25.62 $\pm$  0.15& 25.77 $\pm$  0.14& 25.78 $\pm$  0.17& 26.07 $\pm$  0.17& 26.24 $\pm$  0.21\nl
  V10& 26.01 $\pm$  0.14& 26.26 $\pm$  0.20& 26.29 $\pm$  0.17& 26.23 $\pm$  0.15& 25.68 $\pm$  0.17& 25.61 $\pm$  0.14\nl
  V11& 26.16 $\pm$  0.23& 26.19 $\pm$  0.22& 26.16 $\pm$  0.24& 26.60 $\pm$  0.30& 26.48 $\pm$  0.74& 27.09 $\pm$  0.39\nl
  V12& 25.99 $\pm$  0.18& 25.99 $\pm$  0.19& 26.27 $\pm$  0.24& 26.21 $\pm$  0.17& 26.07 $\pm$  0.20& 25.38 $\pm$  0.12\nl
  V13& 26.64 $\pm$  0.49& 26.07 $\pm$  0.17& 25.80 $\pm$  0.15& 25.87 $\pm$  0.16& 26.11 $\pm$  0.23& 26.27 $\pm$  0.23\nl
  V14& 26.58 $\pm$  0.17& 26.65 $\pm$  0.24& 26.68 $\pm$  0.62& 26.74 $\pm$  0.29&  \nodata         & 25.95 $\pm$  0.14\nl
  V15&  \nodata         & 26.08 $\pm$  0.21& 25.94 $\pm$  0.19& 26.26 $\pm$  0.19& 26.09 $\pm$  0.21& 25.70 $\pm$  0.16\nl
  V16& 26.48 $\pm$  0.37& 26.50 $\pm$  0.23& 25.64 $\pm$  0.16& 25.46 $\pm$  0.36& 26.05 $\pm$  0.20& 26.10 $\pm$  0.22\nl
  V17& 25.50 $\pm$  0.21& 25.63 $\pm$  0.14& 25.77 $\pm$  0.14& 25.88 $\pm$  0.15& 26.02 $\pm$  0.19& 25.50 $\pm$  0.16\nl
  V18& 25.52 $\pm$  0.13& 25.43 $\pm$  0.26& 25.46 $\pm$  0.14& 25.59 $\pm$  0.13& 25.81 $\pm$  0.13& 26.14 $\pm$  0.19\nl
  V19& 25.77 $\pm$  0.15& 25.55 $\pm$  0.36& 25.78 $\pm$  0.43& 26.10 $\pm$  0.37& 26.45 $\pm$  0.22& 26.65 $\pm$  0.24\nl
  V20& 25.83 $\pm$  0.22& 25.99 $\pm$  0.30& 26.21 $\pm$  0.23& 26.12 $\pm$  0.53& 26.64 $\pm$  0.38& 25.43 $\pm$  0.29\nl
  V21& 26.61 $\pm$  0.22& 26.53 $\pm$  0.22& 25.86 $\pm$  0.15& 25.71 $\pm$  0.12& 26.23 $\pm$  0.17& 26.27 $\pm$  0.16\nl
  V22& 26.05 $\pm$  0.15& 26.29 $\pm$  0.18& 26.35 $\pm$  0.17& 26.44 $\pm$  0.27& 25.38 $\pm$  0.13& 25.61 $\pm$  0.37\nl
  V23& 25.91 $\pm$  0.19& 26.25 $\pm$  0.35& 26.12 $\pm$  0.22& 25.19 $\pm$  0.14& 25.42 $\pm$  0.14& 25.70 $\pm$  0.13\nl
  V24& 25.91 $\pm$  0.15& 26.09 $\pm$  0.18& 26.16 $\pm$  0.17& 26.51 $\pm$  0.32& 26.77 $\pm$  0.70& 26.02 $\pm$  0.21\nl
  V25& 26.06 $\pm$  0.18& 26.16 $\pm$  0.22& 26.24 $\pm$  0.23& 26.39 $\pm$  0.23& 25.91 $\pm$  0.18& 25.49 $\pm$  0.17\nl
  V26& 25.29 $\pm$  0.16& 25.48 $\pm$  0.18& 25.56 $\pm$  0.31& 25.75 $\pm$  0.17& 25.86 $\pm$  0.21& 26.33 $\pm$  0.24\nl
  V27& 26.35 $\pm$  0.21& 26.82 $\pm$  0.22& 25.48 $\pm$  0.12& 25.42 $\pm$  0.13& 25.76 $\pm$  0.41& 25.98 $\pm$  0.21\nl
  V28& 26.52 $\pm$  0.18& 26.71 $\pm$  0.21& 26.54 $\pm$  0.15& 26.19 $\pm$  0.20& 26.08 $\pm$  0.21& 26.27 $\pm$  0.21\nl
  V29& 26.19 $\pm$  0.16& 26.44 $\pm$  0.27& 26.27 $\pm$  0.28& 26.39 $\pm$  0.20& 25.40 $\pm$  0.12& 25.63 $\pm$  0.19\nl
  V30& 25.16 $\pm$  0.22& 25.74 $\pm$  0.19& 25.82 $\pm$  0.18& 25.89 $\pm$  0.19& 26.00 $\pm$  0.24& 25.05 $\pm$  0.26\nl
  V31& 25.83 $\pm$  0.13& 26.32 $\pm$  0.20& 26.15 $\pm$  0.15& 26.16 $\pm$  0.16& 26.62 $\pm$  0.26& 25.46 $\pm$  0.14\nl
  V32& 27.03 $\pm$  0.37& 27.46 $\pm$  0.48& 26.99 $\pm$  0.26& 25.95 $\pm$  0.16& 26.49 $\pm$  0.22& 26.61 $\pm$  0.30\nl
  V33& 25.66 $\pm$  0.28& 25.66 $\pm$  0.24& 25.77 $\pm$  0.26& 26.14 $\pm$  0.34& 25.89 $\pm$  0.34& 25.25 $\pm$  0.17\nl
  V34& 25.85 $\pm$  0.21& 26.11 $\pm$  0.23& 26.35 $\pm$  0.23& 26.25 $\pm$  0.21& 25.59 $\pm$  0.17& 25.56 $\pm$  0.29\nl
  V35& 26.84 $\pm$  0.41& 26.79 $\pm$  0.30& 26.55 $\pm$  0.25& 26.90 $\pm$  0.23& 25.95 $\pm$  0.18& 26.30 $\pm$  0.17\nl
  V36& 26.70 $\pm$  0.28& 25.83 $\pm$  0.15& 25.86 $\pm$  0.18& 26.27 $\pm$  0.33& 26.89 $\pm$  0.39& 26.89 $\pm$  0.31\nl
  V37& 25.92 $\pm$  0.16& 26.12 $\pm$  0.23& 26.48 $\pm$  0.21& 26.55 $\pm$  0.49&  \nodata         & 26.30 $\pm$  0.21\nl
  V38& 26.49 $\pm$  0.17& 25.67 $\pm$  0.14& 25.83 $\pm$  0.15& 25.92 $\pm$  0.34& 26.49 $\pm$  0.19& 26.66 $\pm$  0.23\nl
  V39& 26.53 $\pm$  0.21& 25.85 $\pm$  0.15& 25.66 $\pm$  0.14& 26.14 $\pm$  0.20& 26.48 $\pm$  0.23& 26.60 $\pm$  0.20\nl
  V40& 25.68 $\pm$  0.17& 25.92 $\pm$  0.24& 25.94 $\pm$  0.19& 26.22 $\pm$  0.25& 26.30 $\pm$  0.29& 25.54 $\pm$  0.22\nl
  V41& 25.51 $\pm$  0.23& 25.61 $\pm$  0.38& 25.96 $\pm$  0.21& 26.03 $\pm$  0.25& 25.89 $\pm$  0.17& 25.57 $\pm$  0.22\nl
  V42& 26.58 $\pm$  0.24& 25.46 $\pm$  0.14& 25.88 $\pm$  0.16& 26.08 $\pm$  0.14& 26.47 $\pm$  0.20& 26.55 $\pm$  0.22\nl
  V43& 25.63 $\pm$  0.13& 26.12 $\pm$  0.41& 26.20 $\pm$  0.19& 26.54 $\pm$  0.21& 25.97 $\pm$  0.17& 25.39 $\pm$  0.34\nl
  V44& 26.05 $\pm$  0.18& 26.46 $\pm$  0.22& 26.89 $\pm$  0.47& 26.88 $\pm$  0.52& 25.86 $\pm$  0.14& 26.18 $\pm$  0.15\nl
  V45& 26.39 $\pm$  0.22& 26.56 $\pm$  0.20& 26.73 $\pm$  0.24& 27.51 $\pm$  0.50& 26.78 $\pm$  0.25& 26.60 $\pm$  0.20\nl
  V46& 25.72 $\pm$  0.16& 25.97 $\pm$  0.15& 26.19 $\pm$  0.43& 26.72 $\pm$  0.27& 26.64 $\pm$  0.22& 25.81 $\pm$  0.28\nl
  V47& 25.88 $\pm$  0.33& 26.54 $\pm$  0.25& 26.51 $\pm$  0.20& 25.76 $\pm$  0.16& 26.12 $\pm$  0.19& 26.35 $\pm$  0.19\nl
  V48& 26.70 $\pm$  0.23& 27.08 $\pm$  0.38& 26.84 $\pm$  0.28& 26.17 $\pm$  0.15& 26.96 $\pm$  0.32& 26.98 $\pm$  0.29\nl
  V49& 26.94 $\pm$  0.31& 26.21 $\pm$  0.16& 26.58 $\pm$  0.24& 27.40 $\pm$  0.33& 27.28 $\pm$  0.37& 26.31 $\pm$  0.20\nl
  V50& 26.07 $\pm$  0.17& 26.58 $\pm$  0.31& 26.84 $\pm$  0.27& 26.81 $\pm$  0.51& 26.21 $\pm$  0.24& 26.85 $\pm$  0.35\nl
  V51& 26.93 $\pm$  0.29& 27.03 $\pm$  0.29& 26.53 $\pm$  0.17& 26.90 $\pm$  0.21& 27.10 $\pm$  0.40& 26.68 $\pm$  0.24\nl
  V52& 27.20 $\pm$  0.31& 27.39 $\pm$  0.33& 26.96 $\pm$  0.29& 26.51 $\pm$  0.58& 26.84 $\pm$  0.23& 26.61 $\pm$  0.18\nl
\enddata
\end{deluxetable}

\tablenum{5}
\tablewidth{33pc}
\begin{deluxetable}{ccccc}
\tablecaption{I Photometry for NGC 1365 Cepheids}
\tablehead{
\colhead{}&
\multicolumn{4}{c}{Heliocentric Julian Date 2449000.+}\nl
\colhead{Cepheid}&
\colhead{936.297}&
\colhead{944.069}&
\colhead{959.358}&
\colhead{980.403}}
\startdata
   V1& 24.26 $\pm$  0.13& 24.22 $\pm$  0.12& 24.37 $\pm$  0.15& 24.55 $\pm$  0.17\nl
   V2& 24.38 $\pm$  0.13& 24.35 $\pm$  0.13& 23.87 $\pm$  0.12& 24.05 $\pm$  0.13\nl
   V3& 24.30 $\pm$  0.12& 24.23 $\pm$  0.17& 23.87 $\pm$  0.11& 23.79 $\pm$  0.26\nl
   V4& 24.18 $\pm$  0.19& 24.24 $\pm$  0.16& 24.43 $\pm$  0.15& 24.12 $\pm$  0.14\nl
   V5& 24.54 $\pm$  0.31& 24.82 $\pm$  0.16& 25.26 $\pm$  0.16& 24.82 $\pm$  0.15\nl
   V6& 25.34 $\pm$  0.24& 25.25 $\pm$  0.25& 24.95 $\pm$  0.18& 25.29 $\pm$  0.36\nl
   V7& 25.07 $\pm$  0.17& 24.39 $\pm$  0.13& 24.51 $\pm$  0.15& 24.99 $\pm$  0.18\nl
   V8& 24.95 $\pm$  0.15& 24.31 $\pm$  0.14& 24.37 $\pm$  0.20& 24.90 $\pm$  0.22\nl
   V9& 24.84 $\pm$  0.19& 24.97 $\pm$  0.18& 24.65 $\pm$  0.14& 24.95 $\pm$  0.25\nl
  V10& 24.87 $\pm$  0.15& 24.68 $\pm$  0.19& 25.06 $\pm$  0.15& 24.86 $\pm$  0.14\nl
  V11& 25.14 $\pm$  0.30& 25.33 $\pm$  0.23& 24.87 $\pm$  0.18& 25.34 $\pm$  0.25\nl
  V12& 24.93 $\pm$  0.18& 24.58 $\pm$  0.14& 24.95 $\pm$  0.19& 24.99 $\pm$  0.18\nl
  V13& 25.06 $\pm$  0.20& 25.36 $\pm$  0.18& 25.82 $\pm$  0.30& 25.28 $\pm$  0.25\nl
  V14& 25.50 $\pm$  0.22& 25.59 $\pm$  0.18& 25.39 $\pm$  0.20& 25.04 $\pm$  0.18\nl
  V15& 25.01 $\pm$  0.22& 25.37 $\pm$  0.21& 24.64 $\pm$  0.14& 24.99 $\pm$  0.24\nl
  V16& 24.63 $\pm$  0.21& 24.46 $\pm$  0.33& 25.13 $\pm$  0.20& 24.73 $\pm$  0.15\nl
  V17& 24.92 $\pm$  0.23& 25.15 $\pm$  0.17& 24.34 $\pm$  0.20& 24.83 $\pm$  0.14\nl
  V18& 24.83 $\pm$  0.16& 25.01 $\pm$  0.16& 25.15 $\pm$  0.17& 24.97 $\pm$  0.15\nl
  V19& 25.01 $\pm$  0.16& 25.28 $\pm$  0.19& 24.94 $\pm$  0.17& 25.28 $\pm$  0.21\nl
  V20& 25.42 $\pm$  0.19& 25.32 $\pm$  0.21& 25.03 $\pm$  0.15& 25.08 $\pm$  0.54\nl
  V21& 24.95 $\pm$  0.15& 25.15 $\pm$  0.17& 24.68 $\pm$  0.88& 24.63 $\pm$  0.36\nl
  V22& 25.42 $\pm$  0.22& 24.92 $\pm$  0.15& 24.87 $\pm$  0.14& 24.75 $\pm$  0.14\nl
  V23& 25.16 $\pm$  0.21& 24.56 $\pm$  0.18& 25.01 $\pm$  0.20& 24.68 $\pm$  0.16\nl
  V24& 25.29 $\pm$  0.19& 25.52 $\pm$  0.24& 24.98 $\pm$  0.15& 25.68 $\pm$  0.22\nl
  V25& 25.04 $\pm$  0.20& 25.21 $\pm$  0.29& 24.88 $\pm$  0.16& 25.17 $\pm$  0.18\nl
  V26& 24.61 $\pm$  0.21& 24.66 $\pm$  0.18& 24.81 $\pm$  0.19& 24.54 $\pm$  0.37\nl
  V27& 25.49 $\pm$  0.32& 24.88 $\pm$  0.17& 25.07 $\pm$  0.18& 24.84 $\pm$  0.17\nl
  V28& 25.71 $\pm$  0.25& 25.97 $\pm$  0.28& 25.57 $\pm$  0.16& 25.31 $\pm$  0.24\nl
  V29& 25.11 $\pm$  0.21& 25.53 $\pm$  0.30& 24.72 $\pm$  0.28& 24.76 $\pm$  0.17\nl
  V30& 24.69 $\pm$  0.16& 24.97 $\pm$  0.25& 24.60 $\pm$  0.18& 25.19 $\pm$  0.28\nl
  V31& 24.90 $\pm$  0.15& 25.10 $\pm$  0.26& 24.87 $\pm$  0.15& 25.64 $\pm$  0.25\nl
  V32& 26.56 $\pm$  0.48& 26.21 $\pm$  0.44& 25.75 $\pm$  0.20& 25.53 $\pm$  0.20\nl
  V33& 24.53 $\pm$  0.43& 24.76 $\pm$  0.28& 24.58 $\pm$  0.22& 24.98 $\pm$  0.45\nl
  V34& 25.23 $\pm$  0.24& 25.30 $\pm$  0.25& 24.86 $\pm$  0.23& 25.00 $\pm$  0.31\nl
  V35& 25.66 $\pm$  0.24& 26.01 $\pm$  0.28& 25.35 $\pm$  0.22& 25.21 $\pm$  0.22\nl
  V36& 25.77 $\pm$  0.34& 25.35 $\pm$  0.39& 25.85 $\pm$  0.34& 25.62 $\pm$  0.29\nl
  V37& 25.13 $\pm$  0.18& 25.42 $\pm$  0.23& 25.18 $\pm$  0.17& \nodata          \nl
  V38& 25.79 $\pm$  0.30& 25.13 $\pm$  0.17& 25.71 $\pm$  0.18& 25.38 $\pm$  0.19\nl
  V39& 25.69 $\pm$  0.29& 24.87 $\pm$  0.15& 25.73 $\pm$  0.29& 25.30 $\pm$  0.19\nl
  V40& 24.72 $\pm$  0.24& 24.90 $\pm$  0.26& 25.26 $\pm$  0.28& 25.24 $\pm$  0.29\nl
  V41& 25.27 $\pm$  0.21& 24.90 $\pm$  0.17& 25.24 $\pm$  0.20& 25.56 $\pm$  0.32\nl
  V42& 25.72 $\pm$  0.25& 25.11 $\pm$  0.43& 25.67 $\pm$  0.24& 25.64 $\pm$  0.27\nl
  V43& 25.33 $\pm$  0.19& 25.09 $\pm$  0.41& 25.14 $\pm$  0.16& 25.47 $\pm$  0.18\nl
  V44& 25.73 $\pm$  0.28& 25.44 $\pm$  0.22& 25.25 $\pm$  0.17& 25.51 $\pm$  0.23\nl
  V45& 26.04 $\pm$  0.38& 25.89 $\pm$  0.28& 26.18 $\pm$  0.30& 25.92 $\pm$  0.31\nl
  V46& 25.40 $\pm$  0.19& 24.98 $\pm$  0.17& 25.81 $\pm$  0.29& 25.65 $\pm$  0.18\nl
  V47& 24.87 $\pm$  0.17& 25.07 $\pm$  0.26& 25.01 $\pm$  0.15& 24.96 $\pm$  0.18\nl
  V48& 25.58 $\pm$  0.18& 25.72 $\pm$  0.70& 25.44 $\pm$  0.49& 25.82 $\pm$  0.25\nl
  V49& 25.55 $\pm$  0.25& 26.29 $\pm$  0.37& 25.95 $\pm$  0.29& 25.89 $\pm$  0.23\nl
  V50& 26.19 $\pm$  0.60& 26.01 $\pm$  0.44& 26.08 $\pm$  0.52& 25.74 $\pm$  0.37\nl
  V51& 26.36 $\pm$  0.42& 25.90 $\pm$  0.29& 26.00 $\pm$  0.26& 26.07 $\pm$  0.27\nl
  V52&  \nodata         & 25.91 $\pm$  0.25& 25.87 $\pm$  0.29& 26.37 $\pm$  0.39\nl
\enddata
\end{deluxetable}

\begin{deluxetable}{cccccccccccc}
\tablenum{6}
\tablewidth{40pc}
\tablecaption{ALLFRAME Photomeric Parameters for NGC 1365 Cepheids}
\tablehead{
\colhead{ID}&
\colhead{$V_{int}$}& \colhead{$\sigma$}& 
\colhead{$V_{ph}$}& \colhead{$\sigma$}&
\colhead{$I_{int}$}& \colhead{$\sigma$} & 
\colhead{$I_{ph}$}& \colhead{$\sigma$} &
\colhead{$(V-I)_{int}$}& 
\colhead{$(V-I)_{ph}$}& \colhead{PL}}
\startdata
V1&   25.62 & 0.07& 25.66 &  0.07& 24.41 & 0.07 & 24.40 & 0.07 &1.21& 1.25& \nl
V2&   24.88 & 0.07& 25.00 &  0.08& 24.07 & 0.12 & 24.16 & 0.11 &0.80& 0.84& \nl
V3&   24.85 & 0.09& 24.96 &  0.10& 23.96 & 0.12 & 24.03 & 0.11 &0.89& 0.93& \nl
V4&   25.55 & 0.06& 25.52 &  0.06& 24.32 & 0.07 & 24.29 & 0.06 &1.22& 1.23& \nl
V5&   26.22 & 0.09& 26.21 &  0.09& 25.03 & 0.15 & 24.98 & 0.14 &1.20& 1.23& \nl
V6&   26.05 & 0.09& 26.08 &  0.09& 25.10 & 0.09 & 25.19 & 0.08 &0.95& 0.90& +\nl
V7&   25.67 & 0.13& 25.67 &  0.13& 24.71 & 0.15 & 24.70 & 0.15 &0.96& 0.97& +\nl
V8&   25.68 & 0.11& 25.70 &  0.11& 24.64 & 0.15 & 24.60 & 0.15 &1.04& 1.10& \nl
V9&   25.87 & 0.10& 25.91 &  0.11& 24.80 & 0.07 & 24.83 & 0.06 &1.07& 1.08& +\nl
V10&  25.97 & 0.08& 26.01 &  0.08& 24.98 & 0.09 & 24.93 & 0.08 &0.99& 1.08& +\nl
V11&  26.55 & 0.12& 26.60 &  0.12& 25.17 & 0.10 & 25.14 & 0.10 &1.38& 1.46& +\nl
V12&  25.96 & 0.08& 25.99 &  0.08& 24.89 & 0.09 & 24.90 & 0.09 &1.06& 1.09& +\nl
V13&  26.33 & 0.10& 26.34 &  0.10& 25.45 & 0.14 & 25.52 & 0.15 &0.88& 0.82& +\nl
V14&  26.47 & 0.10& 26.53 &  0.10& 25.43 & 0.11 & 25.46 & 0.11 &1.04& 1.07& +\nl
V15&  25.95 & 0.08& 25.99 &  0.08& 24.96 & 0.13 & 24.96 & 0.13 &0.99& 1.03& +\nl
V16&  26.10 & 0.09& 26.13 &  0.09& 24.84 & 0.13 & 24.87 & 0.14 &1.27& 1.26& +\nl
V17&  25.67 & 0.09& 25.71 &  0.10& 24.72 & 0.15 & 24.78 & 0.15 &0.95& 0.94& +\nl
V18&  25.86 & 0.09& 25.84 &  0.09& 25.07 & 0.07 & 25.03 & 0.06 &0.79& 0.81& +\nl
V19&  26.25 & 0.12& 26.23 &  0.12& 25.17 & 0.08 & 25.11 & 0.08 &1.08& 1.11& +\nl
V20&  26.10 & 0.12& 26.15 &  0.13& 25.03 & 0.12 & 25.22 & 0.08 &1.07& 0.92& +\nl
V21&  26.25 & 0.09& 26.27 &  0.09& 24.97 & 0.12 & 24.85 & 0.10 &1.27& 1.42& +\nl
V22&  25.99 & 0.09& 26.03 &  0.10& 25.09 & 0.14 & 25.06 & 0.13 &0.90& 0.97& +\nl
V23&  25.83 & 0.10& 25.84 &  0.10& 24.92 & 0.12 & 24.95 & 0.13 &0.92& 0.88& +\nl
V24&  26.20 & 0.12& 26.26 &  0.12& 25.28 & 0.14 & 25.36 & 0.13 &0.92& 0.90& \nl
V25&  25.99 & 0.11& 26.06 &  0.12& 25.01 & 0.07 & 25.07 & 0.06 &0.98& 0.98& +\nl
V26&  25.90 & 0.11& 25.87 &  0.11& 24.77 & 0.08 & 24.69 & 0.05 &1.14& 1.17& +\nl
V27&  26.16 & 0.13& 26.15 &  0.13& 25.12 & 0.13 & 25.14 & 0.13 &1.05& 1.01& +\nl
V28&  26.44 & 0.06& 26.46 &  0.06& 25.65 & 0.12 & 25.68 & 0.12 &0.78& 0.78& \nl
V29&  26.05 & 0.11& 26.09 &  0.11& 25.07 & 0.16 & 25.09 & 0.17 &0.98& 1.00& \nl
V30&  25.61 & 0.11& 25.65 &  0.12& 24.87 & 0.12 & 24.91 & 0.12 &0.74& 0.74& \nl
V31&  26.15 & 0.12& 26.18 &  0.12& 25.13 & 0.15 & 25.21 & 0.16 &1.02& 0.97& +\nl
V32&  26.77 & 0.11& 26.83 &  0.12& 25.94 & 0.20 & 26.10 & 0.20 &0.83& 0.73& \nl
V33&  25.70 & 0.07& 25.75 &  0.07& 24.71 & 0.09 & 24.77 & 0.09 &0.99& 0.98& \nl
V34&  25.89 & 0.08& 25.97 &  0.08& 25.15 & 0.09 & 25.14 & 0.09 &0.75& 0.83& +\nl
V35&  26.57 & 0.11& 26.63 &  0.11& 25.55 & 0.15 & 25.65 & 0.16 &1.01& 0.98& +\nl
V36&  26.54 & 0.11& 26.48 &  0.11& 25.58 & 0.10 & 25.64 & 0.09 &0.96& 0.84& \nl
V37&  26.40 & 0.13& 26.48 &  0.14& 25.50 & 0.16 & 25.30 & 0.08 &0.90& 1.18& \nl
V38&  26.37 & 0.10& 26.31 &  0.10& 25.48 & 0.13 & 25.49 & 0.13 &0.89& 0.82& +\nl
V39&  26.29 & 0.10& 26.25 &  0.10& 25.45 & 0.18 & 25.37 & 0.17 &0.84& 0.87& +\nl
V40&  26.05 & 0.09& 26.09 &  0.09& 25.05 & 0.12 & 25.05 & 0.11 &0.99& 1.04& \nl
V41&  25.91 & 0.08& 25.93 &  0.08& 25.27 & 0.12 & 25.26 & 0.12 &0.63& 0.66& \nl
V42&  26.25 & 0.10& 26.22 &  0.10& 25.54 & 0.12 & 25.46 & 0.13 &0.70& 0.76& +\nl
V43&  26.18 & 0.11& 26.21 &  0.12& 25.31 & 0.08 & 25.26 & 0.07 &0.87& 0.95& +\nl
V44&  26.71 & 0.16& 26.80 &  0.17& 25.72 & 0.15 & 25.58 & 0.10 &0.99& 1.22& +\nl
V45&  26.99 & 0.11& 26.97 &  0.11& 26.01 & 0.06 & 26.00 & 0.06 &0.97& 0.97& +\nl
V46&  26.36 & 0.12& 26.36 &  0.12& 25.44 & 0.16 & 25.36 & 0.16 &0.92& 1.00& +\nl
V47&  26.17 & 0.09& 26.23 &  0.09& 25.02 & 0.04 & 25.00 & 0.04 &1.15& 1.23& +\nl
V48&  26.69 & 0.12& 26.74 &  0.12& 25.67 & 0.07 & 25.65 & 0.07 &1.03& 1.09& +\nl
V49&  26.87 & 0.11& 26.85 &  0.11& 25.84 & 0.14 & 25.84 & 0.14 &1.03& 1.01& +\nl
V50&  26.69 & 0.08& 26.68 &  0.08& 26.07 & 0.09 & 26.05 & 0.09 &0.62& 0.63& \nl
V51&  26.90 & 0.07& 26.96 &  0.08& 26.09 & 0.09 & 26.17 & 0.10 &0.81& 0.79& +\nl
V52&  26.99 & 0.12& 27.06 &  0.13& 26.27 & 0.18 & 26.16 & 0.15 &0.73& 0.90& \nl
\enddata
\end{deluxetable}

\begin{deluxetable}{ccccccc}
\tablenum{7}
\tablewidth{30pc}
\tablecaption{Cepheid Distance Moduli to NGC 1365}
\tablehead{
\colhead{Dataset}&
\colhead{Chip}&\colhead{No.} & \colhead{$\mu_{V}$}&
\colhead{$\mu_{I}$} & \colhead{A$_{V}$} &
\colhead{$\mu_{0}$}}
\startdata
\multicolumn{7}{c}{ALLFRAME Photometry}\nl
phase&1 &       9& 31.68& 31.53& 0.367& 31.31\nl
phase&2 &      12& 31.72& 31.55& 0.408& 31.31\nl
phase&3 &       8& 31.72& 31.54& 0.426& 31.29\nl
phase&4 &       5& 31.71& 31.55& 0.382& 31.33\nl
phase&1-4 &    34& 31.70& 31.54& 0.398& 31.31\nl
phase&1-4 &    52& 31.68& 31.54& 0.331& 31.35\nl
phase&1,3,4 &  22& 31.70& 31.54& 0.392& 31.31\nl
&&&&\nl
intensity&1&      9& 31.66& 31.52& 0.332& 31.32\nl
intensity&2&     12& 31.70& 31.56& 0.349& 31.36\nl
intensity&3&      8& 31.69& 31.56& 0.322& 31.37\nl
intensity&4&      5& 31.67& 31.52& 0.367& 31.31\nl
intensity&1-4&   34& 31.68& 31.55& 0.341& 31.34\nl
intensity&1-4&   52& 31.65& 31.54& 0.266& 31.38\nl
intensity&1,3,4& 22& 31.67& 31.54& 0.336& 31.34\nl
&&&\nl
\multicolumn{7}{c}{DoPHOT Photometry}\nl
phase&1&       9& 31.63& 31.52& 0.280& 31.35\nl
phase&2&      12& 31.56& 31.36& 0.500& 31.06\nl
phase&3&       8& 31.79& 31.63& 0.397& 31.39\nl
phase&4&       5& 31.59& 31.50& 0.237& 31.36\nl
phase&1-4&    34& 31.64& 31.49& 0.379& 31.26\nl
phase&1-4&    52& 31.58& 31.46& 0.299& 31.28\nl
phase&1,3,4&  22& 31.68& 31.55& 0.313& 31.37\nl
&&\nl
intensity&1&      9& 31.63& 31.51& 0.285& 31.34\nl
intensity&2&     12& 31.57& 31.37& 0.478& 31.09\nl
intensity&3&      8& 31.77& 31.62& 0.366& 31.41\nl
intensity&4&      5& 31.59& 31.48& 0.260& 31.33\nl
intensity&1-4&   34& 31.64& 31.48& 0.369& 31.27\nl
intensity&1-4&   52& 31.58& 31.46& 0.297& 31.28\nl
intensity&1,3,4& 22& 31.67& 31.55& 0.309& 31.36\nl
\enddata
\end{deluxetable}

\begin{deluxetable}{lc}
\tablenum{8}
\tablewidth{30pc}
\tablecaption{NGC 1365 Distance Modulus Error Budget}
\tablehead{
\colhead{Source of Uncertainty}&
\colhead{Error}}
\startdata
\multicolumn{2}{l}{\bf Cepheid Period Luminosity Relation Calibration}\nl
(1) LMC True Distance Modulus & $\pm$ 0.10 \nl
(2) $V$-band PL Zero Point & $\pm$ 0.05 \nl
(3) $I$-band PL Zero Point & $\pm$ 0.03 \nl
(S1) PL Systematic Uncertainty & $\pm$ 0.12 \nl 
&\nl
\multicolumn{2}{l}{\bf NGC 1365 Distance Modulus}\nl
(4) HST WFPC2 $V$-band Photometry & $\pm$ 0.07 \nl
(5) HST WFPC2 $I$-band Photometry & $\pm$ 0.06 \nl
(R1) Cepheid True Modulus & $\pm$ 0.18 \nl
&\nl
\multicolumn{2}{l}{\bf NGC 1365 Cepheid PL Relation Fitting}\nl
(6) $V$-band PL Fitting & $\pm$ 0.05 \nl
(7) $I$-band PL Fitting & $\pm$ 0.06 \nl
(R2) Cepheid True Modulus & $\pm$ 0.08 \nl
&\nl
(S2) {\bf Metallicity Uncertainty} & $\pm$ 0.12\nl
&\nl
(S3) {\bf Long vs Short Zero Point} & $\pm$ 0.05\nl
&\nl
{\bf Total Uncertainty} & \nl
(R) Random Errors & $\pm$ 0.20 \nl
(S) Systematic Error& $\pm$ 0.18 \nl
\enddata
\end{deluxetable}

\tablenum{A1}
\tablewidth{40pc}
\begin{deluxetable}{ccccccccc}
\tablecaption{Secondary Standard Stars in NGC 1365}
\tablehead{
\colhead{Chip-ID}&
\colhead{X}&
\colhead{Y}&
\colhead{RA (J2000)}&
\colhead{Dec (J2000)}&
\colhead{$V$}&
\colhead{$\sigma$}&
\colhead{$I$}&
\colhead{$\sigma$}}
\startdata
  1-97 & 495.9& 782.1& 3:33:42.43& -36:09:02.8& 24.28 & 0.10 & 23.13 & 0.10\nl
 1-101 & 515.3& 592.1& 3:33:42.84& -36:09:09.9& 24.05 & 0.10 & 23.34 & 0.10\nl
  1-47 & 235.8& 583.8& 3:33:43.66& -36:09:02.0& 23.38 & 0.10 & 23.13 & 0.10\nl
  1-78 & 313.2& 555.3& 3:33:43.50& -36:09:05.3& 23.92 & 0.10 & 23.12 & 0.10\nl
  1-48 & 675.4& 549.1& 3:33:42.48& -36:09:16.2& 23.59 & 0.10 & 22.70 & 0.10\nl
  1-83 & 778.9& 412.7& 3:33:42.52& -36:09:23.9& 23.94 & 0.10 & 23.61 & 0.10\nl
 1-148 & 726.8& 372.2& 3:33:42.76& -36:09:23.8& 24.37 & 0.10 & 23.52 & 0.10\nl
   1-5 & 123.0& 290.2& 3:33:44.69& -36:09:08.9& 22.27 & 0.10 & 21.14 & 0.10\nl
  1-92 & 563.1& 288.7& 3:33:43.44& -36:09:21.9& 23.89 & 0.10 & 23.93 & 0.11\nl
 1-105 & 702.2& 251.3& 3:33:43.13& -36:09:27.3& 24.12 & 0.10 & 23.82 & 0.11\nl
 1-120 & 620.3& 224.8& 3:33:43.43& -36:09:25.8& 24.78 & 0.10 & 22.97 & 0.10\nl
 1-147 & 638.9& 192.8& 3:33:43.45& -36:09:27.4& 24.24 & 0.10 & 24.09 & 0.11\nl
&&&&&&\nl
 2-204 &  91.9&  96.4& 3:33:45.72& -36:09:07.2& 23.77 & 0.10 & 22.00 & 0.10\nl
 2-659 & 123.7& 230.4& 3:33:46.38& -36:08:56.1& 24.19 & 0.10 & 24.21 & 0.11\nl
  2-62 & 373.4& 319.8& 3:33:45.60& -36:08:31.4& 22.95 & 0.10 & 21.74 & 0.11\nl
 2-333 & 511.4& 324.3& 3:33:44.88& -36:08:20.7& 23.68 & 0.10 & 22.90 & 0.10\nl
 2-276 & 462.7& 362.9& 3:33:45.39& -36:08:21.9& 23.42 & 0.10 & 23.21 & 0.10\nl
 2-467 & 255.3& 505.2& 3:33:47.39& -36:08:28.3& 23.97 & 0.10 & 23.42 & 0.10\nl
 2-457 & 273.4& 531.8& 3:33:47.45& -36:08:25.2& 24.09 & 0.10 & 23.18 & 0.10\nl
 2-288 & 244.6& 553.5& 3:33:47.74& -36:08:25.9& 23.39 & 0.10 & 23.16 & 0.10\nl
 2-588 & 525.7& 572.3& 3:33:46.35& -36:08:03.5& 24.31 & 0.10 & 23.86 & 0.10\nl
 2-319 & 522.2& 601.6& 3:33:46.55& -36:08:01.9& 23.47 & 0.10 & 23.27 & 0.10\nl
 2-654 & 594.2& 623.4& 3:33:46.30& -36:07:55.0& 24.31 & 0.10 & 24.09 & 0.10\nl
 2-138 & 596.1& 643.8& 3:33:46.42& -36:07:53.6& 23.69 & 0.10 & 21.64 & 0.10\nl
 2-281 & 477.0& 675.6& 3:33:47.26& -36:08:00.5& 23.67 & 0.10 & 22.72 & 0.10\nl
&&&&&&\nl
  3-73 & 338.3&  81.9& 3:33:47.59& -36:08:57.7& 24.57 & 0.10 & 24.30 & 0.10\nl
  3-21 & 706.0& 173.5& 3:33:50.38& -36:08:40.9& 23.95 & 0.10 & 22.94 & 0.10\nl
   3-4 & 453.7& 201.5& 3:33:48.95& -36:08:59.3& 22.22 & 0.10 & 19.92 & 0.10\nl
  3-15 & 318.8& 481.1& 3:33:49.60& -36:09:29.2& 23.70 & 0.10 & 23.50 & 0.10\nl
  3-72 & 304.7& 499.6& 3:33:49.61& -36:09:31.5& 24.71 & 0.10 & 23.96 & 0.11\nl
  3-18 & 567.8& 547.9& 3:33:51.51& -36:09:18.1& 23.69 & 0.10 & 23.63 & 0.10\nl
  3-64 & 481.1& 556.0& 3:33:51.02& -36:09:24.4& 24.51 & 0.10 & 24.25 & 0.11\nl
  3-11 & 311.1& 568.8& 3:33:50.02& -36:09:36.3& 23.68 & 0.10 & 22.27 & 0.10\nl
  3-37 & 446.6& 580.6& 3:33:50.93& -36:09:28.5& 24.23 & 0.10 & 24.29 & 0.11\nl
   3-1 & 449.8& 691.4& 3:33:51.54& -36:09:36.6& 20.95 & 0.10 & 19.60 & 0.10\nl
   3-2 & 719.9& 715.5& 3:33:53.36& -36:09:21.0& 21.77 & 0.10 & 19.13 & 0.10\nl
   3-3 & 125.8& 768.6& 3:33:49.93& -36:10:03.4& 21.91 & 0.10 & 19.52 & 0.10\nl
&&&&&&\nl
  4-52 & 355.4&  94.0& 3:33:46.85& -36:09:41.4& 23.39 & 0.10 & 22.53 & 0.10\nl
 4-288 & 665.5& 100.4& 3:33:48.44& -36:10:05.5& 24.39 & 0.10 & 24.47 & 0.11\nl
 4-322 & 208.0& 106.1& 3:33:45.99& -36:09:30.9& 24.55 & 0.10 & 24.13 & 0.10\nl
 4-359 & 249.2& 126.3& 3:33:46.08& -36:09:35.4& 24.69 & 0.10 & 24.62 & 0.11\nl
 4-173 & 304.7& 140.1& 3:33:46.29& -36:09:40.5& 24.18 & 0.10 & 23.45 & 0.10\nl
 4-105 & 410.9& 218.9& 3:33:46.35& -36:09:53.7& 23.69 & 0.10 & 23.72 & 0.11\nl
 4-140 & 329.5& 233.4& 3:33:45.83& -36:09:48.4& 23.84 & 0.10 & 23.55 & 0.10\nl
 4-418 & 446.8& 246.1& 3:33:46.37& -36:09:58.1& 24.74 & 0.10 & 24.74 & 0.11\nl
 4-214 & 370.6& 274.1& 3:33:45.79& -36:09:54.1& 24.58 & 0.10 & 23.10 & 0.10\nl
 4-222 & 474.3& 345.8& 3:33:45.88& -36:10:06.6& 24.23 & 0.10 & 23.86 & 0.10\nl
 4-116 & 454.0& 369.4& 3:33:45.63& -36:10:06.6& 23.63 & 0.10 & 23.46 & 0.10\nl
 4-260 & 472.0& 383.0& 3:33:45.64& -36:10:08.8& 24.28 & 0.10 & 23.93 & 0.10\nl
  4-43 & 221.8& 394.2& 3:33:44.25& -36:09:50.4& 24.44 & 0.10 & 21.73 & 0.10\nl
  4-89 & 285.2& 417.2& 3:33:44.43& -36:09:56.7& 24.50 & 0.10 & 22.32 & 0.10\nl
 4-242 & 331.2& 532.3& 3:33:43.95& -36:10:07.6& 24.38 & 0.10 & 23.57 & 0.10\nl
 4-108 & 348.6& 572.8& 3:33:43.79& -36:10:11.5& 23.92 & 0.10 & 22.89 & 0.10\nl
 4-221 & 616.4& 752.3& 3:33:44.07& -36:10:43.5& 24.14 & 0.10 & 24.11 & 0.11\nl
\enddata
\end{deluxetable}

\begin{deluxetable}{lccccc}
\tablenum{A2}
\tablewidth{20pc}
\tablecaption{Possible Variable Stars in NGC 1365}
\tablehead{
\colhead{chip-id}&
\colhead{x} &
\colhead{y} &
\colhead{V} & 
\colhead{V-I}&
\colhead{P(days)}} 
\startdata
1-1298 & 113 & 735 & 26.12 &1.22 &28.6/29.1\nl
1-1699 & 569 &  76 & 26.54 &0.94 &23.7/25.5\nl
1-2514 & 458 & 618 & 26.37 &0.44 &36.4/41.4\nl
1-1671 & 515 & 683 & 26.81 &1.74 &35.0\nl
1-2931 & 524 & 511 & 27.08 &1.14 &17.2/18.2\nl
1-2420 & 169 & 670 & 26.92 &1.25 &27.9/29.1\nl
1-1085 & 763 & 184 & 26.19 &0.92 &16.8\nl
1-1070 & 379 & 208 & 25.86 &1.76 &30.3\nl
1-0066 & 262 & 322 & 27.60 &0.88 &36.3\nl
&&\nl
2-1375 & 709 & 230 & 25.51 &1.33 &41.1\nl
2-4325 & 549 & 491 & 26.57 &1.09 &21.0\nl
2-4581 & 254 & 570 & 27.05 &0.78 &35.3\nl
2-3175 & 760 & 268 & 26.49 &0.78 &24.2\nl
2-5456 & 322 & 411 & 26.73 &0.84 &13.9\nl
2-4065 & 392 & 555 & 26.56 &0.72 &22.1\nl
2-6655 & 622 & 703 & 26.97 &0.43 &39.3\nl
2-1969 & 679 & 490 & 25.83 &1.00 &24.0\nl
&\nl
3-2056 & 335 & 120 & 27.04 &1.02 &22.1/23.5\nl
3-1696 & 495 & 444 & 26.91 &1.18 &29.0/33.3\nl
3-776  & 465 & 575 & 26.36 &1.05 &24.0/29.0\nl
3-1650 &  79 & 374 & 26.73 &0.92 &27.1/21.8\nl
3-3233 & 502 & 408 & 27.16 &1.11 &19.8/36.3\nl
3-946  &  78 & 173 & 26.37 &0.36 &12.3/24.5\nl
3-2855 & 399 & 318 & 27.09 &0.37 &22.0\nl
3-3747 & 468 & 446 & 27.31 &-0.69 &20.0\nl
3-2658 & 478 & 465 & 26.88 &0.43 &13.8/23.0\nl
&\nl
4-1766 & 121 & 412 & 26.50 &1.37 &24.3\nl
4-3052 & 515 & 551 & 26.56 &0.65 &14.2\nl
4-1523 & 181 & 661 & 26.01 &0.64 &36.5\nl
4-3271 & 132 & 670 & 26.79 &0.66 &39.0\nl
\enddata
\end{deluxetable}



\begin{figure}
\figurenum{2}
\plotfiddle{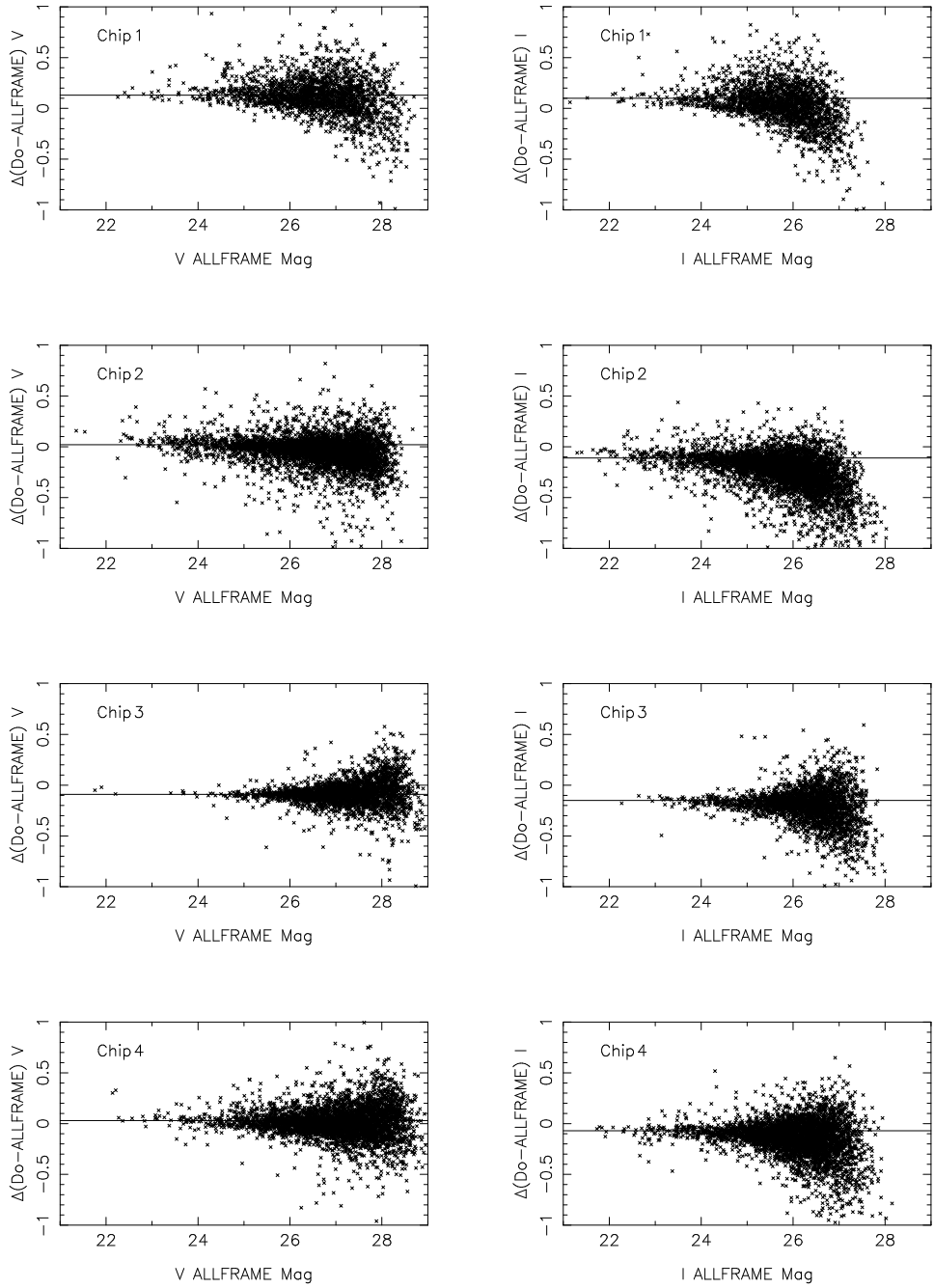}{6in}{0}{70}{70}{-200}{0}
\caption{Comparison of the ALLFRAME and DoPHOT \ngc{1365}
photometry. Table 2 lists the mean differences between the photometry.}
\end{figure}

\newpage

\begin{figure}
\figurenum{3}
\plotfiddle{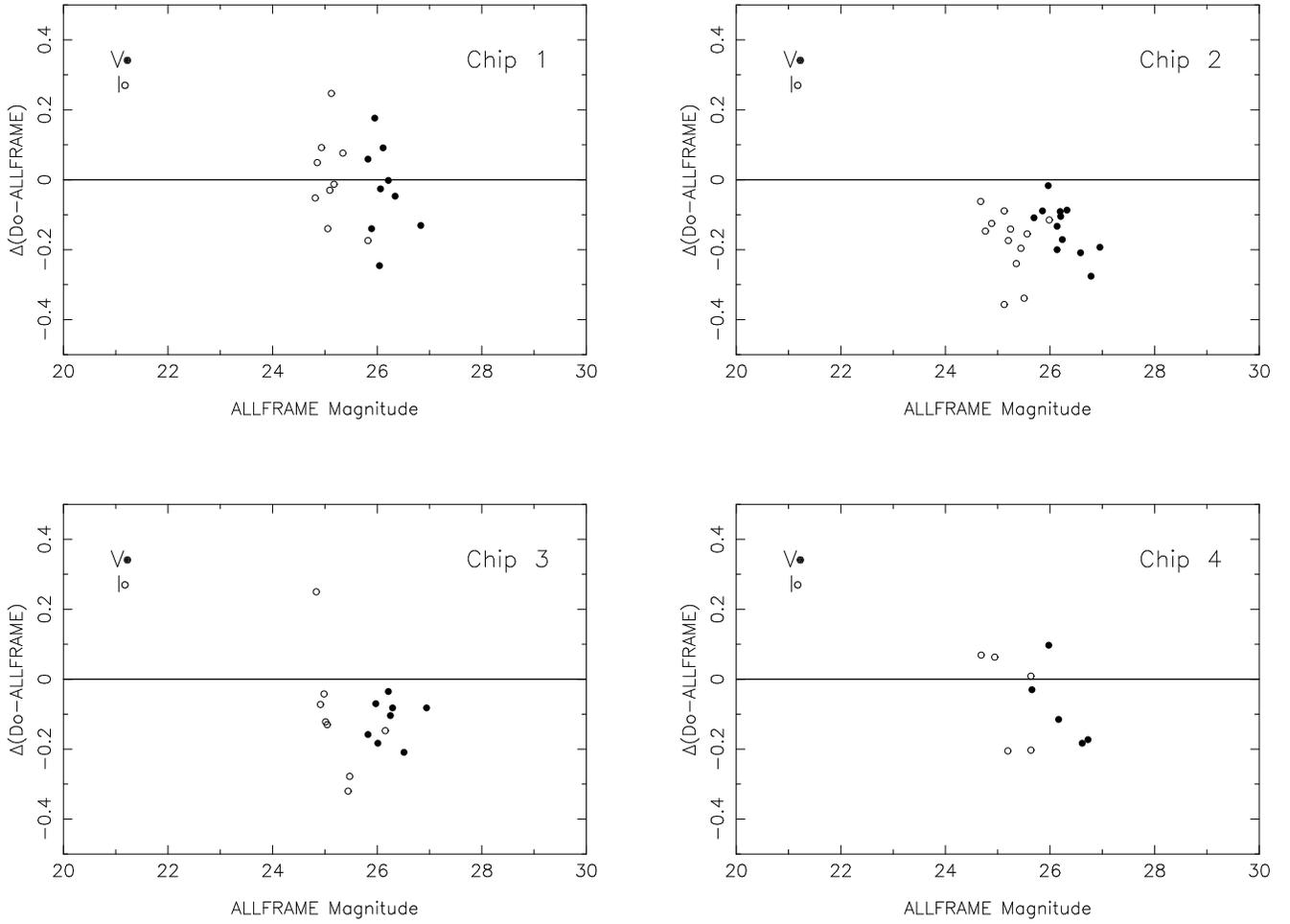}{6in}{270}{70}{70}{-270}{450}
\caption{Comparison of the ALLFRAME and DoPHOT
photometry for the 34 Cepheids in \ngc{1365}. The 
filled circles are the $V$ comparisons and the open circles are
the $I$ comparisons. Table 2 lists the mean differences between
the ALLFRAME and DoPHOT photometry.}
\end{figure}

\newpage








\newpage

\begin{figure}
\figurenum{6}
\plotfiddle{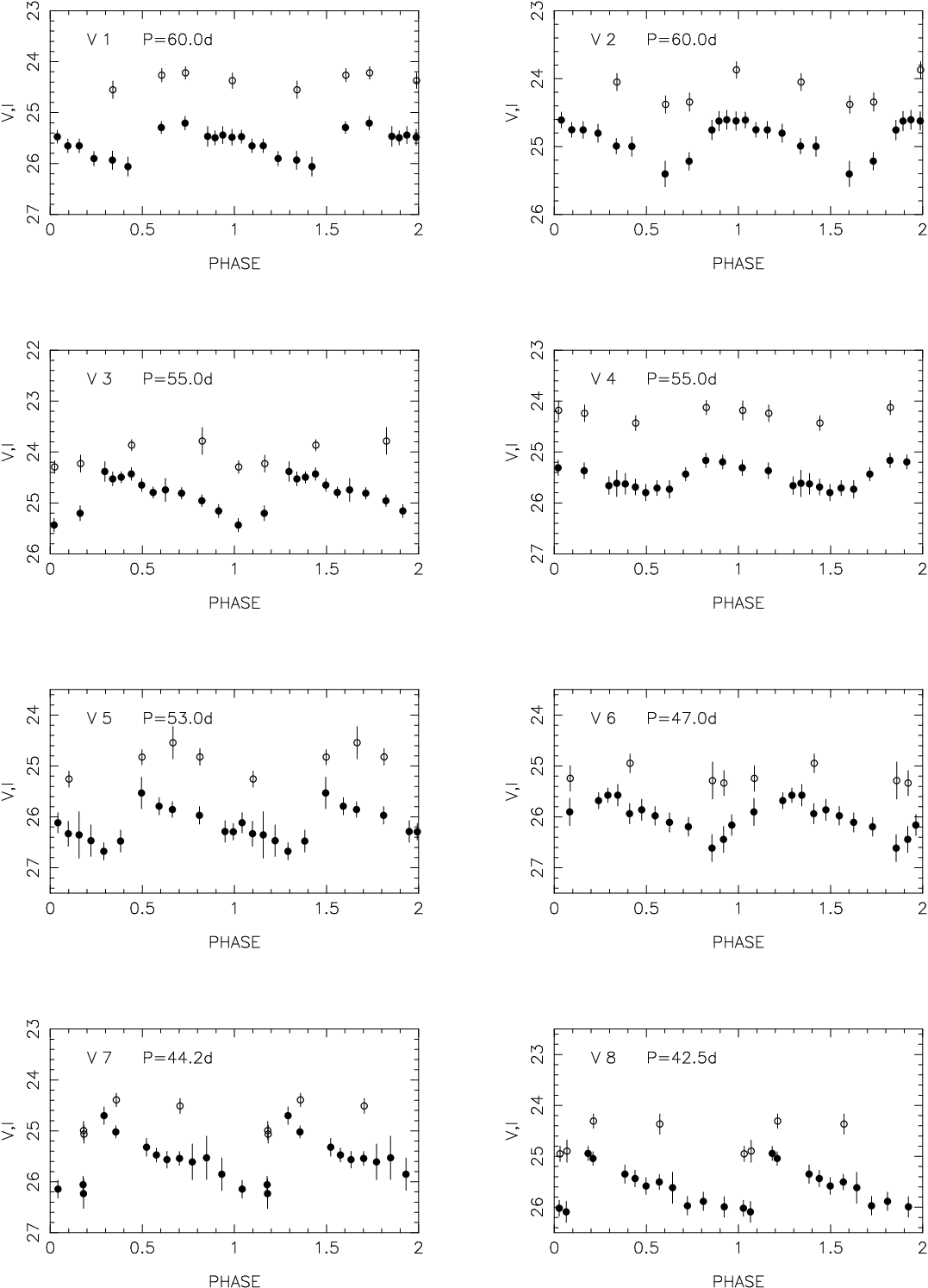}{6in}{0}{70}{70}{-230}{0}
\caption{$V$ and $I$ light curves are presented for the 52 
Cepheids in \ngc{1365}. The data are repeated over a second cycle for clarity.
$V$ data are filled circles. $I$ data are open circles.}
\end{figure}

\newpage

\begin{figure}
\figurenum{6}
\plotfiddle{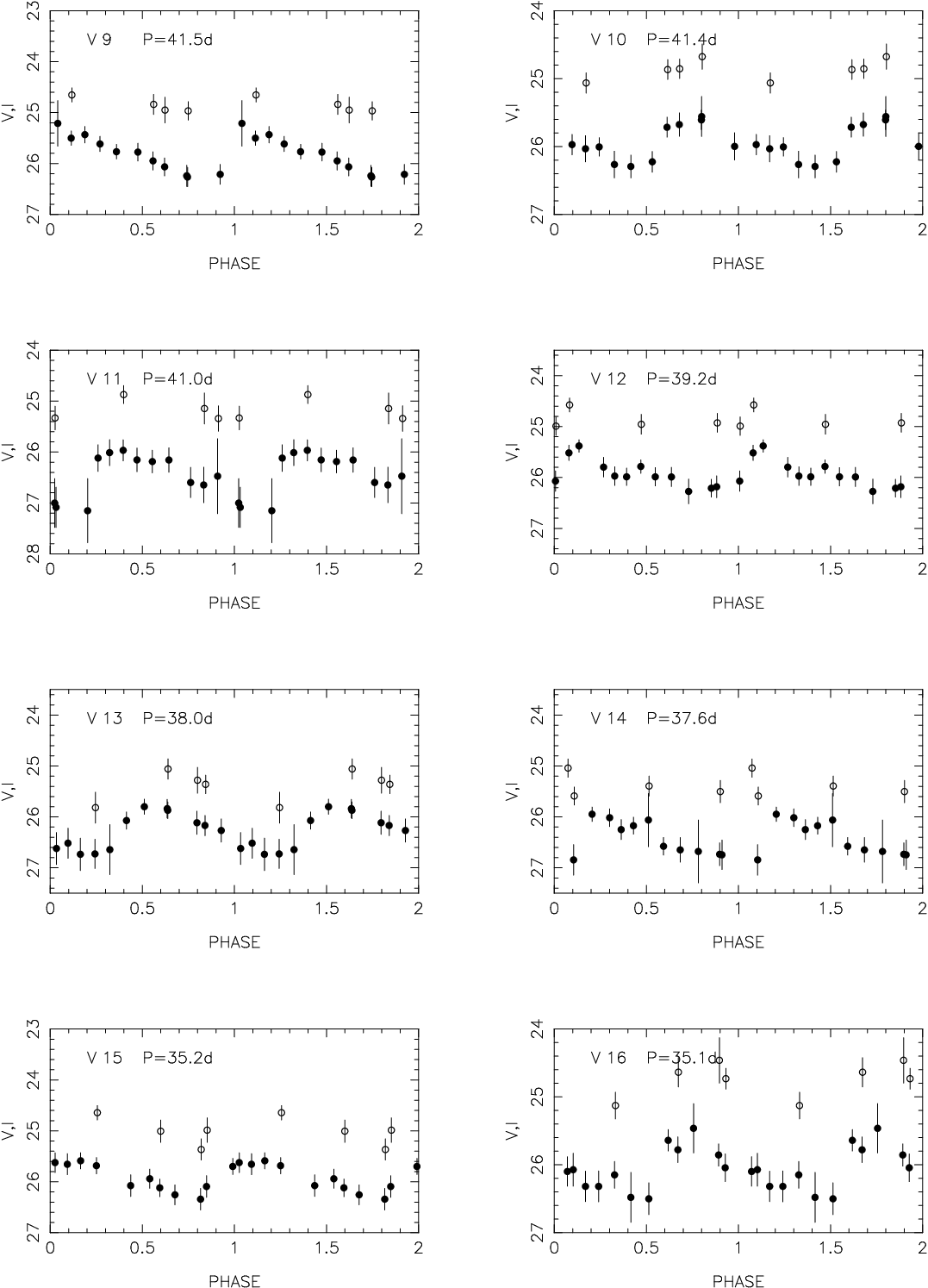}{6in}{0}{70}{70}{-230}{0}
\caption{$V$ and $I$ light curves are presented for the 52
Cepheids in \ngc{1365}. The data are repeated over a second cycle for clarity.
$V$ data are filled circles. $I$ data are open circles.}
\end{figure}

\newpage

\begin{figure}
\figurenum{6}
\plotfiddle{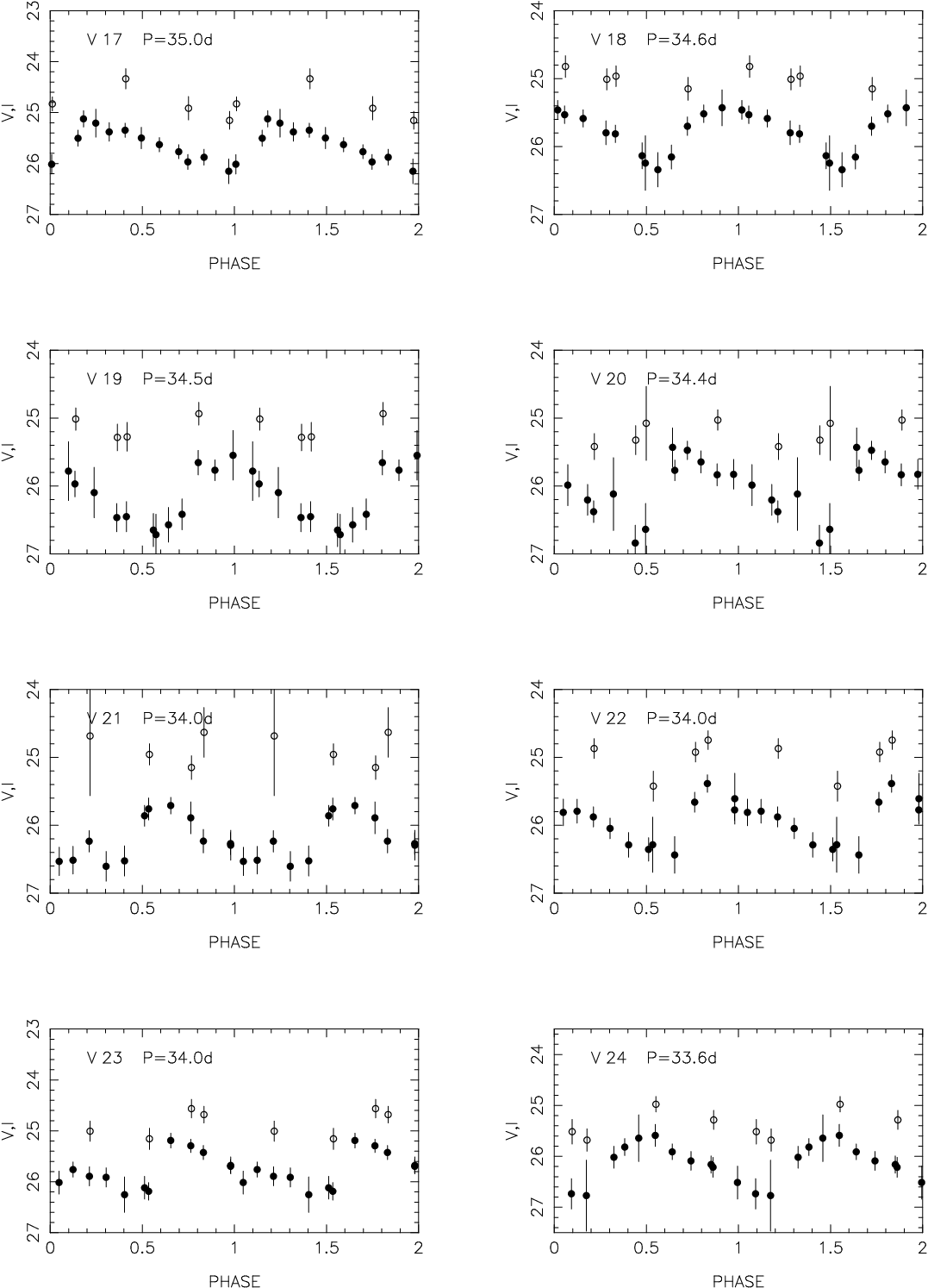}{6in}{0}{70}{70}{-230}{0}
\caption{$V$ and $I$ light curves are presented for the 52
Cepheids in \ngc{1365}. The data are repeated over a second cycle for clarity.
$V$ data are filled circles. $I$ data are open circles.}
\end{figure}

\newpage

\begin{figure}
\figurenum{6}
\plotfiddle{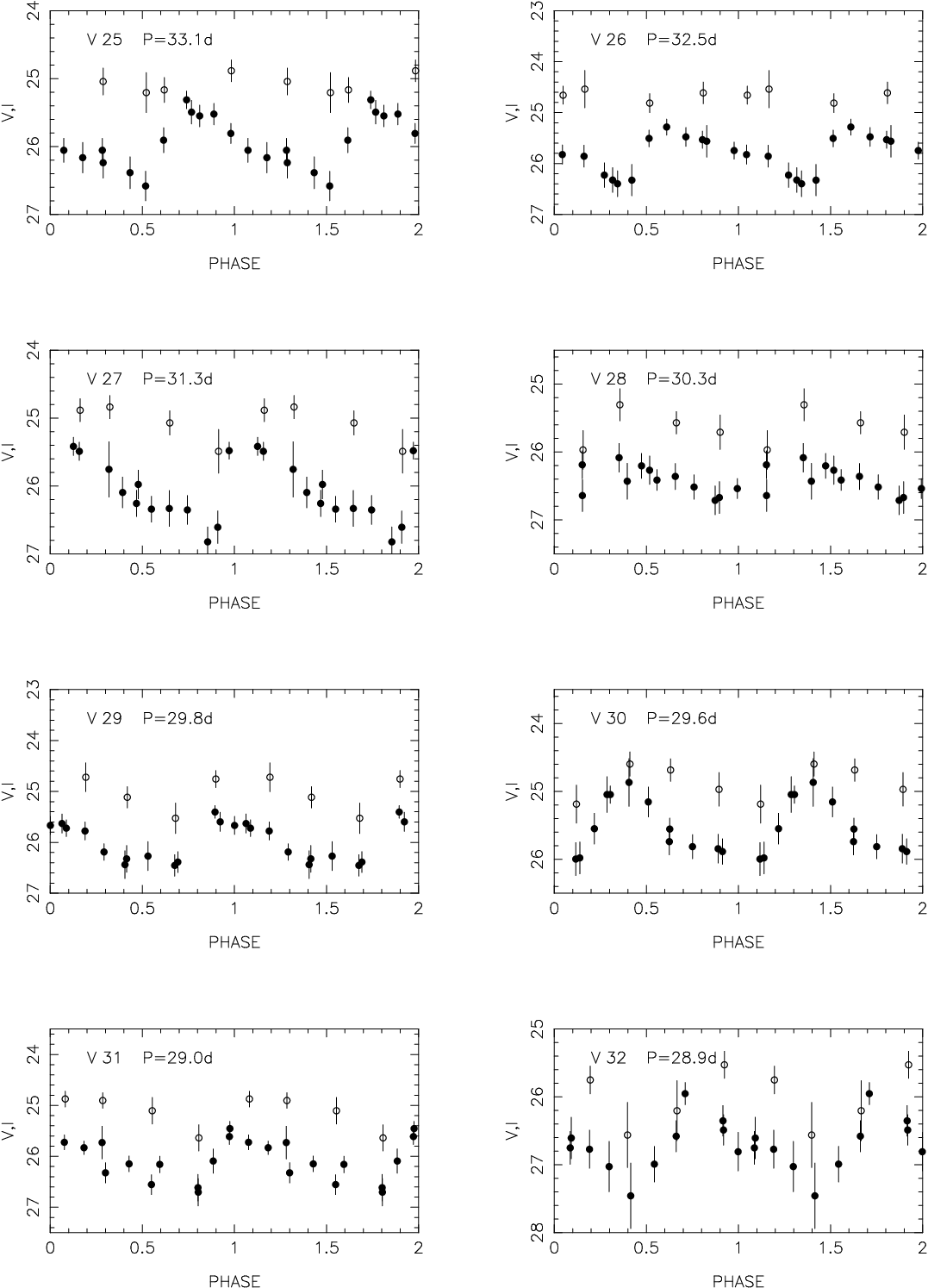}{6in}{0}{70}{70}{-230}{0}
\caption{$V$ and $I$ light curves are presented for the 52
Cepheids in \ngc{1365}. The data are repeated over a second cycle for clarity.
$V$ data are filled circles. $I$ data are open circles.}
\end{figure}

\newpage

\begin{figure}
\figurenum{6}
\plotfiddle{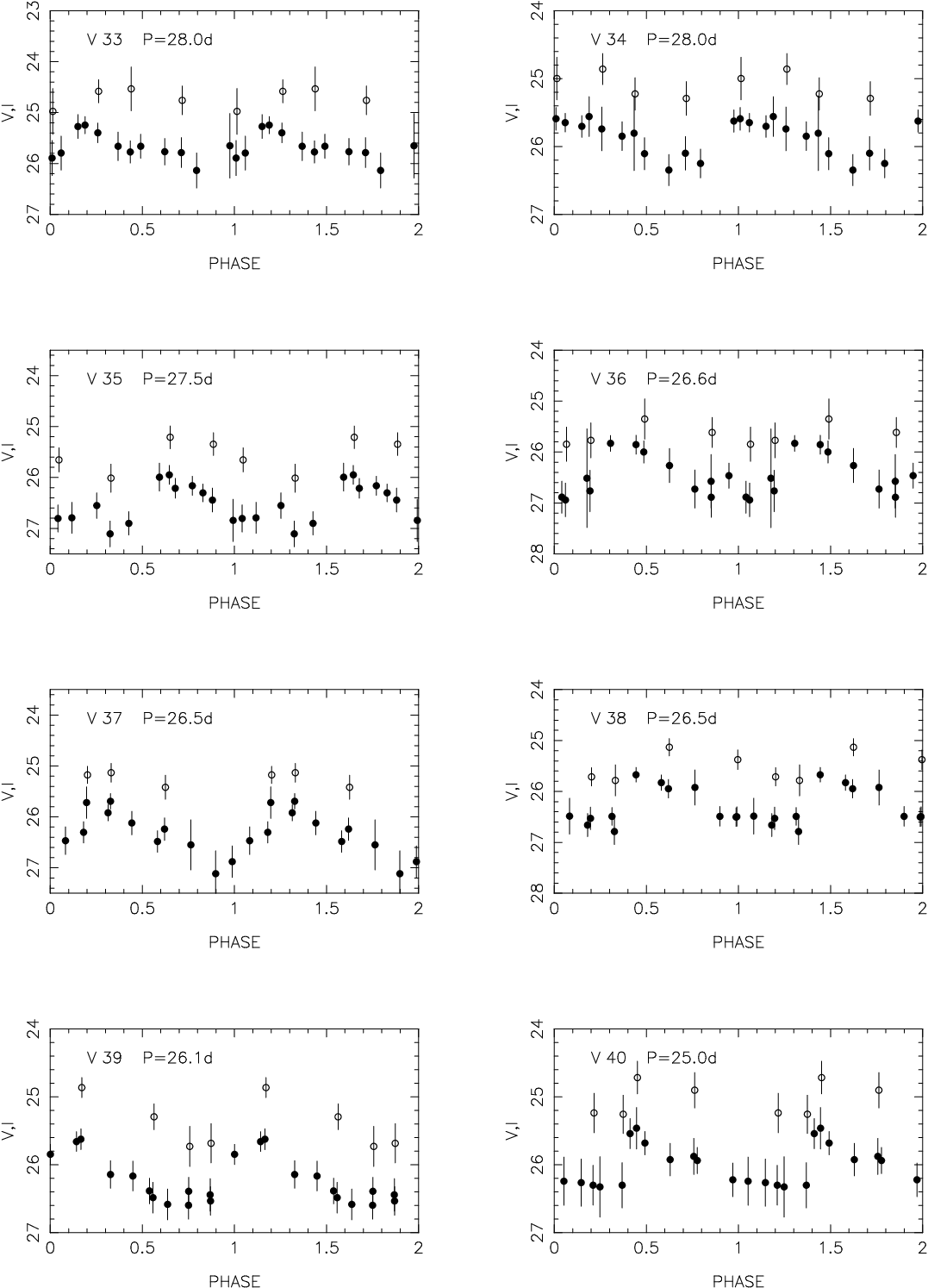}{6in}{0}{70}{70}{-230}{0}
\caption{$V$ and $I$ light curves are presented for the 52
Cepheids in \ngc{1365}. The data are repeated over a second cycle for clarity.
$V$ data are filled circles. $I$ data are open circles.}
\end{figure}

\newpage

\begin{figure}
\figurenum{6}
\plotfiddle{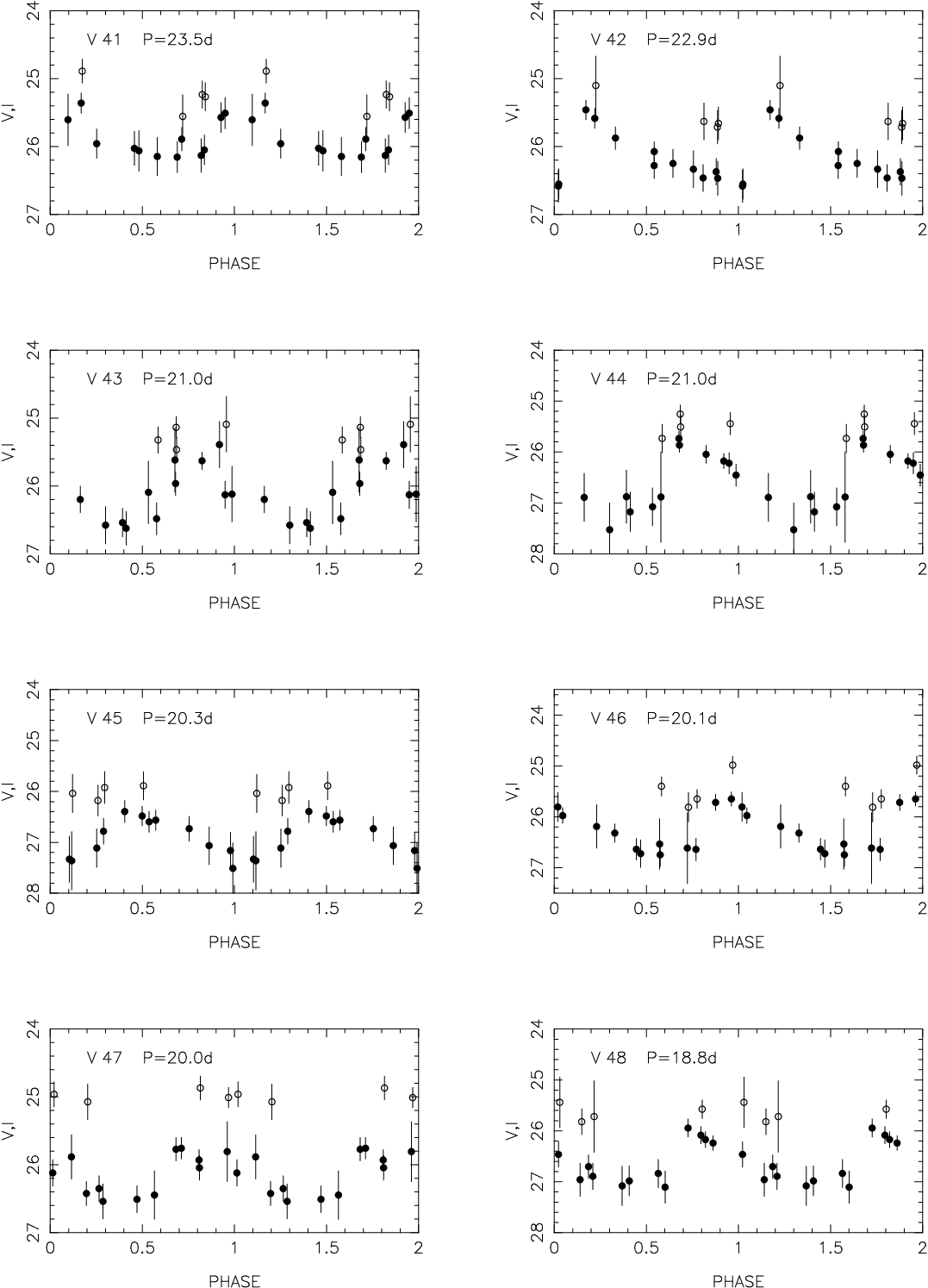}{6in}{0}{70}{70}{-230}{0}
\caption{$V$ and $I$ light curves are presented for the 52
Cepheids in \ngc{1365}. The data are repeated over a second cycle for clarity.
$V$ data are filled circles. $I$ data are open circles.}
\end{figure}

\newpage

\begin{figure}
\figurenum{6}
\plotfiddle{f6g.ps}{6in}{0}{70}{70}{-230}{0}
\caption{$V$ and $I$ light curves are presented for the 52
Cepheids in \ngc{1365}. The data are repeated over a second cycle for clarity.
$V$ data are filled circles. $I$ data are open circles.}
\end{figure}

\newpage

\begin{figure}
\figurenum{7}
\plotfiddle{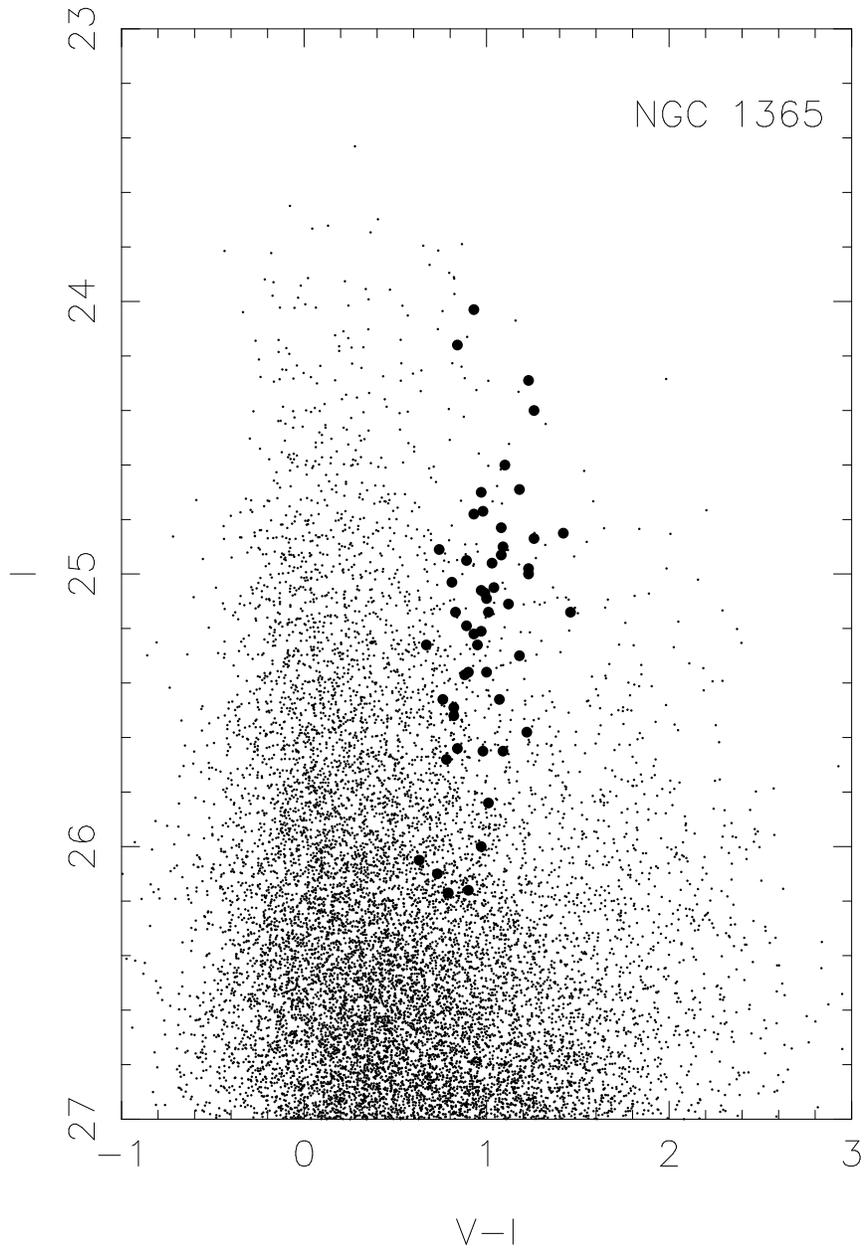}{6in}{0}{70}{70}{-200}{0}
\caption{Color-magnitude diagram for the HST
field of \ngc{1365}. Cepheids are indicated by the large filled circles.}
\end{figure}

\newpage

\begin{figure}
\figurenum{8}
\plotfiddle{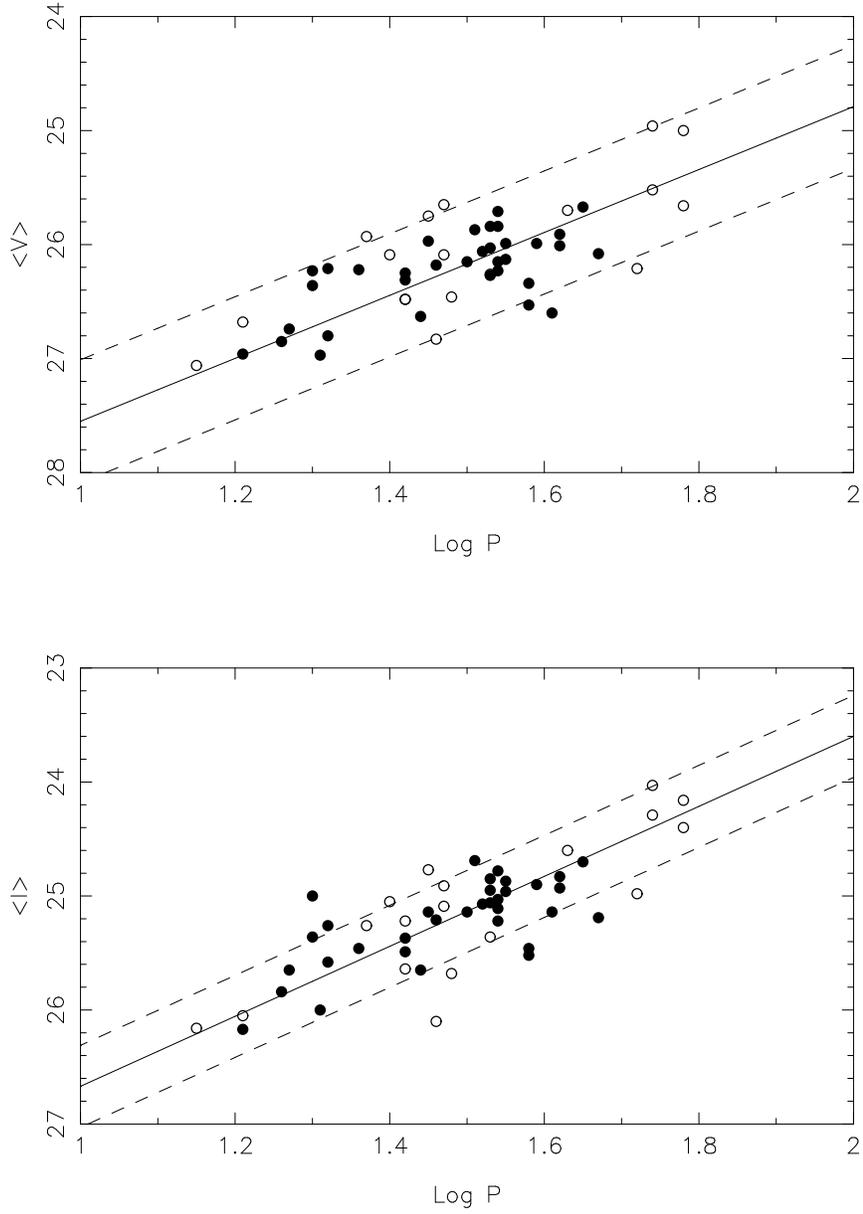}{6in}{0}{65}{65}{-200}{0}
\caption{ALLFRAME $V$ and $I$ period-luminosity relations 
for the Cepheids in \ngc{1365}.  The filled circles are those Cepheids used to 
determine the distance to \ngc{1365}. The open circles are the rest of the 
Cepheids. The solid line is the best fit to the \ngc{1365} data.
The dotted lines indicate the expected scatter due to the intrinsic
width of the Cepheid instability strip.}
\end{figure}
\end{document}